\documentclass{aastex}
\usepackage{emulateapj5}
\begin{document}

\newcommand{\hi}{$h^{-1}$~}
\newcommand{\kms}{~km~s$^{-1}$}
\newcommand{\logh}{+5log$h$}

\title{The Nature of E+A Galaxies in Intermediate Redshift
  Clusters$^{1,2}$}

\footnotetext[1]{Based on observations with the NASA/ESA Hubble Space
  Telescope, obtained at the Space Telescope Science Institute, which
  is operated by the Association of Universities for Research in
  Astronomy, Inc., under NASA contract NAS 5-26555.}
\footnotetext[2]{Based on observations obtained at the W. M. Keck
  Observatory, which is operated jointly by the California Institute of
  Technology and the University of California.}

\author{Kim-Vy H. Tran\footnotemark[3]}
\footnotetext[3]{Current address:  Institute for
  Astronomy, ETH H\"onggerberg, CH-8093 Z\"urich, Switzerland}
\affil{Department of Astronomy \& Astrophysics, University of
California, Santa Cruz, CA 95064}
\email{vy@phys.ethz.ch}

\author{Marijn Franx}
\affil{Leiden Observatory, P.O. Box 9513, 2300 RA Leiden, The
Netherlands}
\email{franx@strw.leidenuniv.nl}

\author{Garth Illingworth}
\affil{University of California Observatories/Lick Observatory,
University of California, Santa Cruz, CA 95064}
\email{gdi@ucolick.org}

\author{Daniel D. Kelson}
\affil{Observatories of the Carnegie Institution of Washington, 813
  Santa Barbara Street, Pasadena, CA, 91101}
\email{kelson@ociw.edu}

\author{Pieter van Dokkum}
\affil{Department of Astronomy, Yale University, New Haven, CT 06520-8101}
\email{dokkum@astro.yale.edu}

\setcounter{footnote}{3}

\begin{abstract}
  
  Combining HST/WFPC2 mosaics with extensive ground-based spectroscopy,
  we study the nature of E+A galaxies in three intermediate redshift
  clusters ($z=0.33,~0.58,~\&~0.83$).  From a sample of $\sim500$
  confirmed cluster members, we isolate 46 E+A candidates to determine
  the E+A fraction and study their physical properties.  Spectral types
  are assigned using Balmer (H$\delta$, H$\gamma$, H$\beta$) and
  [OII]~$\lambda3727$ equivalent widths.  For all members, we have
  galaxy colors, luminosities, Hubble types, and quantitative
  structural parameters.  We also include measured internal velocity
  dispersions for 120 cluster members, and estimate velocity
  dispersions for the rest of the cluster sample using the Fundamental
  Plane.  We find E+A's comprise a non-negligible component
  ($\sim7-13$\%) of the cluster population at these redshifts, and
  their diverse nature indicates a heterogeneous parent population.
  While cluster E+A's are predominantly disk-dominated systems, they
  span the range in Hubble type and bulge-to-total fraction to include
  even early-type members.  Cluster E+A's also cover a wide range in
  luminosity ($L_B\sim0.2-2.5L_B^{\ast}$), internal velocity dispersion
  ($\sigma\sim30-220$\kms), and half-light radius
  ($r_{1/2}\sim0.4-4.3$\hi kpc).  From their velocity dispersions and
  half-light radii, we infer that the descendants of E+A's in our
  highest redshift cluster are massive early-type galaxies.  In
  contrast to the wide range of luminosity and internal velocity
  dispersion spanned by E+A's at higher redshift, only low mass E+A's
  are found in nearby clusters, $e.g.$ Coma.  The observed decrease in
  the characteristic E+A mass is similar to the decrease in luminosity
  of rapidly star-forming field galaxies since $z\sim1$, $i.e.$ galaxy
  ``down-sizing.''  In addition, we argue our statistics imply that
  $\gtrsim30$\% of the E-S0 members have undergone an E+A phase; the
  true fraction could be 100\% if the effects of E+A down-sizing, an
  increasing E+A fraction with redshift, and the conversion of spirals
  into early-types are also considered.  Thus, the E+A phase may indeed
  be an important stage in the transformation of star-forming galaxies
  into early-type members.

\end{abstract}



\section{Introduction}

Post-starburst galaxies \citep[``E+A'';][]{dressler:83} in clusters may
provide the crucial link in the morphological transformation of spiral
galaxies into the elliptical/S0 systems that dominate the cluster
population at the current epoch \citep[cf.][]{butcher:78a}.
Characterized by strong Balmer absorption and little or no
[OII]$\lambda3727$ emission, E+A's also are referred to as k+a/a+k
\citep[hereafter D99]{franx:93,poggianti:99,dressler:99} or
H$\delta$-strong galaxies \citep{couch:87,couch:94}.  Despite the short
window of visibility of the post-starburst phase \citep[$<1.5$
Gyr;][]{couch:87,barger:96,leonardi:96}, E+A members can contribute up
to 20\% of the total cluster population (D99), and have been found in
virtually every spectroscopic cluster survey from $0<z<0.8$
\citep{dressler:83,couch:87,wirth:94,caldwell:96,balogh:99,dressler:99,poggianti:99,tran:02}.

However, the fraction of E+A galaxies in intermediate redshift clusters
($0.2<z<0.8$) is still debated.  The MORPHS survey estimated a high E+A
fraction of $\sim20$\% (D99) while the CNOC1 survey found a much lower
fraction of $\sim2$\% \citep[hereafter B99]{balogh:99}.  These results
conflict as to whether the {\it majority} of cluster members undergo an
E+A phase, or if only a small fraction do.  If the real E+A fraction in
clusters is high, it can be an important constraint on how members have
evolved.  For example, the transformation of late-type spirals via a
post-starburst episode may explain the high number of passive S0
galaxies seen in nearby clusters \citep{dressler:83,couch:87}.

Once a robust E+A sample in intermediate redshift clusters is
established, more fundamental issues such as the nature of the parent
population can be addressed.  Past studies find that while E+A's are
heterogeneous in morphology, they tend to have disks
\citep[D99]{wirth:94,couch:94,caldwell:97,couch:98}.  Rotation
velocities determined for a small subset of cluster E+A's at
$z\lesssim0.3$ also show most of them are rotationally supported
systems \citep[hereafter K00b]{franx:93,caldwell:96,kelson:00b}.
However, it has been suggested that many field E+A's and some cluster
E+A's are the result of processes that would easily disrupt a disk,
e.g. merging and/or strong galaxy-galaxy interactions
\citep{belloni:95,liu:95,zabludoff:96}.  Whether cluster E+A's include
even the most massive galaxies, as suggested by \citet{wirth:94},
remains an open question.

A more challenging but extremely interesting question is what the
descendants of cluster E+A's are.  To answer this question, diagnostics
that are not likely to evolve strongly with redshift are needed, e.g.
internal velocity dispersions ($\sigma$) and half-light radii
($r_{1/2}$).  However, published work on E+A's that utilizes both
$\sigma$ and $r_{1/2}$ have been limited to clusters at $z\lesssim0.3$
\citep[hereafter K00c]{franx:93,caldwell:96,kelson:97,kelson:00c}.  By
applying a similar analysis to E+A's at higher redshifts
($z\gtrsim0.3$), we can determine if even the most massive galaxies in
nearby clusters had an E+A phase in their past.  Only with a
combination of deep spectroscopy and high resolution imaging can both
these parameters be measured with confidence in intermediate redshift
clusters.

In the field, \citet{cowie:96} find the maximum luminosity of galaxies
undergoing rapid star formation has decreased smoothly since $z\sim1$.
This evolution in the characteristic mass of the star-forming
population, referred to as ``down-sizing,'' reflects how mass
assembled, i.e. more massive galaxies formed at higher redshift.
Recent work in clusters suggest down-sizing also can play an important
role in rich environments \citep{poggianti:01,kodama:01,poggianti:03}.
If this is the case, the luminosity and internal velocity dispersion
distributions of cluster E+A's should evolve as a function of
redshift.  By determining if the characteristic mass of cluster
galaxies undergoing the E+A phase evolves, we can provided additional
compelling evidence for down-sizing in clusters.

The high fraction of E+A galaxies in clusters relative to the field
suggests that environment plays a significant role in producing E+A's
(D99).  For example, the E+A phase may be triggered by an external
source such as the strong galaxy-galaxy interactions suggested by
studies of field E+A's \citep{liu:95,zabludoff:96}.  In addition to
direct mergers \citep{vandokkum:99}, the cluster environment provides
numerous other external forces that could trigger an E+A phase, e.g.
ram-pressure stripping \citep{gunn:72}, gas compression
\citep{dressler:83}, perturbation by the tidal field \citep{byrd:90},
and galaxy harassment \citep{moore:96}.  The more disruptive
interactions would create morphological signatures that should be
visible during the E+A lifetime.  Examining the number of E+A's that
show morphological signs of recent mergers or interactions can help
isolate which interactions trigger the E+A phase in clusters.  Also,
by comparing their spatial distribution to cluster substructure, we
can test if the global cluster environment is effective at producing
E+A's, or if they are better correlated with local over-densities.

To address these issues, we must first overcome the inherent difficulty
of isolating E+A galaxies.  A statistically representative sample of
members ($>100$) in {\it each} cluster is needed so that the relatively
few E+A galaxies can be identified.  Membership confirmation and E+A
selection can only be accomplished reliably by obtaining spectra.  In
addition, high resolution imaging is needed at these redshifts
($z>0.3$) to determine physical properties such as structural
parameters and morphological type.  Only by pairing wide-field
HST/WFPC2 imaging with deep ground-based spectroscopy can we adequately
study the E+A galaxies in intermediate redshift clusters.

From extensive spectroscopic surveys of CL1358
\citep[$z=0.33$;][]{fisher:98}, MS2053 \citep[$z=0.58$;][]{tran:02},
and MS1054 \citep[$z=0.83$;][]{tran:99,vandokkum:00,tran:02}, we select
46 E+A candidates from $\sim500$ confirmed cluster members.  Using
HST/WFPC2 mosaics taken of each cluster (all to $R_{BCG}\sim1$\hi Mpc),
we measure the colors, magnitudes, half-light radii, bulge-to-total
fractions, degree of galaxy asymmetry, and morphological type of the
cluster members.  With LRIS \citep{oke:95} on Keck, we also have
measured internal velocity dispersions for 120 cluster members
\citep[K00b]{kelson:97,vandokkum:98b,kelson:01,kelson:03a}.  With these
measured dispersions, accurate colors, and the Fundamental Plane
\citep{faber:87,djorgovski:87}, we estimate velocity dispersions for
the remainder of the sample.  Using the E+A's that satisfy our strict
selection criteria, we determine the E+A fraction in intermediate
redshift clusters, identify characteristics of their parent population,
address what the descendants of these galaxies can be, and discuss the
likely down-sizing of this population.

A brief description of the HST/WFPC2 imaging and ground-based
spectroscopy is provided in \S2.  We describe our E+A selection
criteria and address the discrepancy in the cluster E+A fraction found
by different surveys in \S3.  After identifying the cluster E+A
population, we examine the nature of these systems in \S4.  We discuss
their properties and evolution in \S5, and present our conclusions in
\S6.  Unless otherwise noted, we use $\Omega_M=0.3,
\Omega_{\Lambda}=0.7$, and $H_0=100h$\kms~Mpc$^{-1}$ in this paper.

\section{Summary of Observations and Data}

We select E+A galaxies from a large program studying three X-ray
luminous clusters at $z=0.33,~0.58,~\&~0.83$ (Table~\ref{clusters},
references therein).  Our dataset combines HST/WFPC2 mosaics (each to
$R_{BCG}\sim1$\hi Mpc) of these clusters with extensive ground-based
spectroscopy.  From the spectroscopy, we determine spectral types for
all members \citep{fisher:98,vandokkum:00,tran:02} as well as measure
internal kinematics for a subset
\citep[K00b]{kelson:97,vandokkum:98b,kelson:01,kelson:03a}.  From over
1200 redshifts obtained in the three fields, we isolate $\sim500$
cluster members.  Here we describe briefly the spectra and photometry
used in this paper.


\subsection{Photometry}

The three clusters were imaged by HST/WFPC2 in the F606W and F814W
filters.  The image reduction and photometry are detailed for CL1358,
MS2053, and MS1054 in \citet{vandokkum:98a}, \citet{hoekstra:00}, and
\citet{vandokkum:00} respectively.  Following the method outlined in
\citet{vandokkum:96}, we transform from the WFPC2 filter system to
redshifted Johnson magnitudes using:

\begin{equation}
\begin{array}{l}
B_z = F814W+a(F606W-F814W)+b \\
V_z = F814W+c(F606W-F814W)+d
\end{array}
\end{equation}

\noindent where the constants $\{a,b,c,d\}$ for an E/S0 galaxy
are

\begin{equation}
\begin{array}{ll}
\{1.021, 0.524, 0.204, 0.652\} & z=0.33\\
\{0.354, 0.923, -0.173, 0.959\} & z=0.58\\
\{-0.077, 1.219, -0.524, 1.229\} & z=0.83
\end{array}
\end{equation}

\noindent These apparent magnitudes correspond to integrating the galaxy's
spectral energy distribution through the redshifted Johnson filter
curves.  Apparent magnitude then is converted to an absolute magnitude
adjusted for passive evolution ($M_{Be}$) using the distance moduli in
Table~\ref{clusters}; here we account for simple fading, as determined
from the Fundamental Plane
\citep[$\Delta\log(M/L)\propto-0.40z$;][]{vandokkum:98b}.

In our analysis, we include Hubble types from
\citet{fabricant:00,fabricant:03} who visually typed all members with
$m_{814}<22$.  In the nomenclature adopted by these authors, the
morphological types of \{E, E/S0, S0, S0/Sa, Sa, Sb, Sc, Sd\} were
assigned values of $\{-5, -4, -2, 0, 1, 3, 5, 7\}$; intermediate values
of $\{-3, -1, 2\}$ were also used and mergers assigned a value of 99.
In the following, we consider E-S0 galaxies as having $-5\leq T\leq-1$,
S0/a-Sa galaxies $0\leq T\leq1$, and spirals $2\leq T\leq 15$.

\subsection{Spectroscopy}

In CL1358, spectra of 230 members were obtained at the WHT and MMT.
Details of the target selection and spectral reduction are in
\citet[hereafter F98]{fisher:98}.  Additional high resolution spectra
were taken with LRIS \citep{oke:95} on the Keck Telescope to obtain
internal velocity dispersions of 55 members (K00b).  Keck/LRIS was also
used to obtain redshifts and internal velocity dispersions for both
MS2053 and MS1054.  In MS2053 and MS1054, 150 and 130 members were
confirmed respectively \citep{tran:99,vandokkum:00,tran:02}, and
internal velocity dispersions for 29 and 26 members measured
\citep{kelson:97,vandokkum:98b,kelson:03a}.  Detailed explanations of
the spectral reductions including the wavelength calibration, sky
subtraction, and removal of telluric absorption can be found in the
noted references.

The wavelength coverage for virtually all members in the three clusters
includes [OII]~$\lambda$3727, the 4000~\AA~break, and Balmer lines
(H$\delta$, H$\gamma$, \& H$\beta$).  The inclusion of these spectral
features are of paramount importance as E+A galaxies are defined
spectroscopically as having strong Balmer absorption and no
[OII]~$\lambda3727$ emission
\citep[D99]{dressler:83,zabludoff:96,balogh:99,poggianti:99}.  The
bandpasses used to determine the equivalent widths of these features
are listed in Table~\ref{indices}.


\section{Defining E+A Galaxies}

\subsection{E+A Selection Criteria}

\begin{figure}
\epsscale{0.6}
\plotone{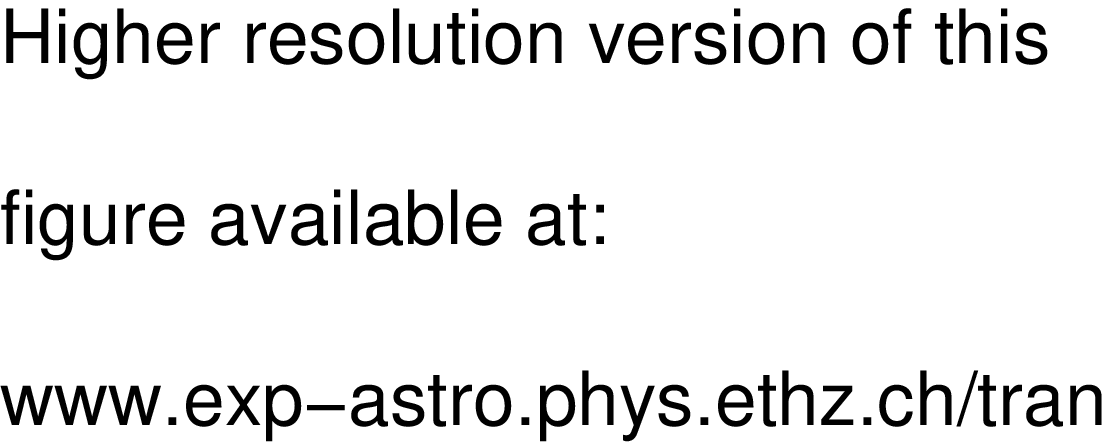}
\caption{\footnotesize Color images made with WFPC2 F814W and F606W imaging of the
  46 E+A candidates from our cluster sample.  The images are
  $10''\times10''$ (CL1358, $z=0.33$), $7''\times7''$ (MS2053,
  $z=0.58$), and $5''\times5''$ (MS1054, $z=0.83$). Members are ordered
  by decreasing $M_{Be}$, and the 14 E+A's above our magnitude and
  $S/N$ cut are noted with an asterix.  For comparison, a
  representative bright elliptical member is included for each cluster:
  they are 1358-256, 2053-1993, and 1054-5450.  (F606W-F814W)
  corresponds closely to redshifted $(B-V)$ and $(U-B)$ for CL1358 and
  MS1054 respectively; it falls between the two colors for MS2053.
\label{ea_images}}
\end{figure}

We select E+A galaxies from the three clusters in our sample
(Table~\ref{clusters}) as having $($H$\delta+$H$\gamma)/2\geq4$~\AA~and
no [OII] emission ($>-5$~\AA).  Although using all three Balmer lines
to select E+A's is the most robust approach \citep{newberry:90},
H$\beta$ is severely compromised by sky lines in MS1054 and so we do
not include H$\beta$ in our selection criteria (see
Table~\ref{ea_sel}).  Due to the wavelength coverage, we cannot
determine the equivalent width of [OII] $\lambda3727$ for $\sim7$\% and
$\sim4$\% of members in MS2053 and MS1054 respectively; for these
galaxies, only H$\delta$ and H$\gamma$ are used to determine their E+A
status.  This adds four E+A's (H3549, H2345, H1746, H408) to the MS2053
sample.  The 46 cluster members that satisfy these criteria are shown
in Fig.~\ref{ea_images} and their physical characteristics are listed
in Table~\ref{ea_sample}.



\begin{figure}
\epsscale{0.6}
\plotone{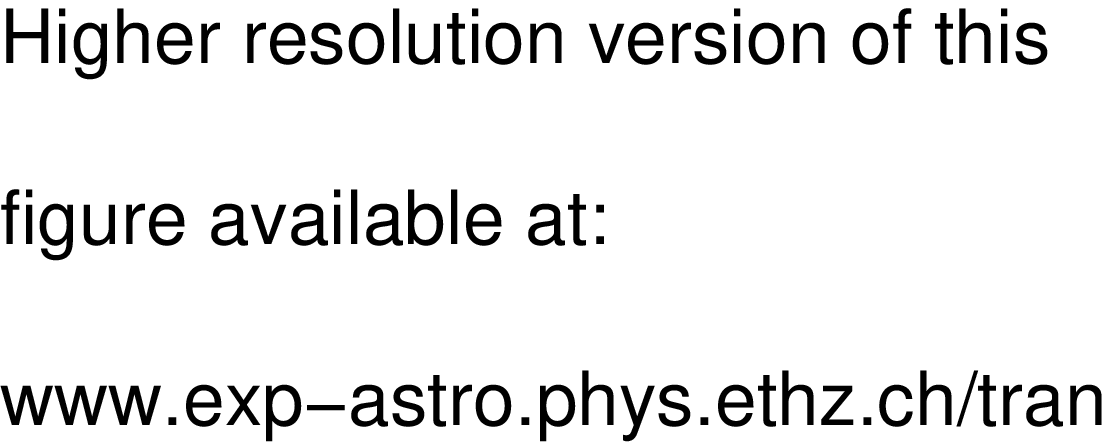}
\caption{Smoothed spectra of the 14 E+A galaxies that satisfy our stringent
  selection criteria.  The vertical dotted lines denote the bandpass
  for [OII]~$\lambda3727$, H$\delta$, H$\gamma$, and H$\beta$.  Note
  the strong absorption of all three Balmer lines for E+A's in CL1358
  and MS2053; H$\beta$ measurements in MS1054 are severely compromised
  by sky lines.  For 1054-6567, lack of OII~$\lambda3727$ emission is
  confirmed from a lower S/N spectrum.
\label{ea_spectra}}
\end{figure}

Because the spectral quality varies for the three clusters, we define a
magnitude limit ($M_{Be}\leq-19.1$\logh, $m_{814}\sim21.9$) set by
MS1054, the highest redshift cluster in our sample.  For uniformity, we
consider only members brighter than this cut when comparing the E+A
population between the clusters.  In selecting E+A's, we also apply a
signal to noise cut ($S/N\geq20$) on the H$\delta$ and H$\gamma$
fluxes.  Only 14 of the 46 E+A candidates satisfy these strict
selection criteria; these spectra are shown in Fig.~\ref{ea_spectra}.


An important point is that the spectroscopic criteria used to identify
E+A galaxies vary depending on author, as can be seen in
Table~\ref{ea_sel}.  The use of different thresholds and lines, $e.g.$
only H$\delta$ versus a combination of Balmer lines, can produce
different E+A fractions even within the same sample.  The most robust
approach is to use all three Balmer lines \citep[cf.][]{newberry:90}
and [OII]$\lambda3727$ but, as noted earlier, this is not possible for
our entire sample.  Thus we restrict ourselves to using
[OII]$\lambda3727$ in combination with H$\delta$ and H$\gamma$.  The
only exceptions in our sample are 2053--3549 and 2053--2345.  Although
we do not have OII EQW's for these two E+A candidates, both have strong
H$\beta$ absorption such that (H$\delta+$H$\gamma+$H$\beta)/3>4$\AA,
and all three Balmer lines have $S/N>20$.  It is very unlikely that
these two E+A candidates also have strong OII emission because in the
CL1358 and MS2053 samples, $>90$\% of members with
(H$\delta+$H$\gamma+$H$\beta)/3>4$\AA~show no significant
[OII]$\lambda3727$ emission.  We confirm 1054--6567's lack of OII
emission from a lower $S/N$ spectrum.

\begin{figure}
\epsscale{1}
\plotone{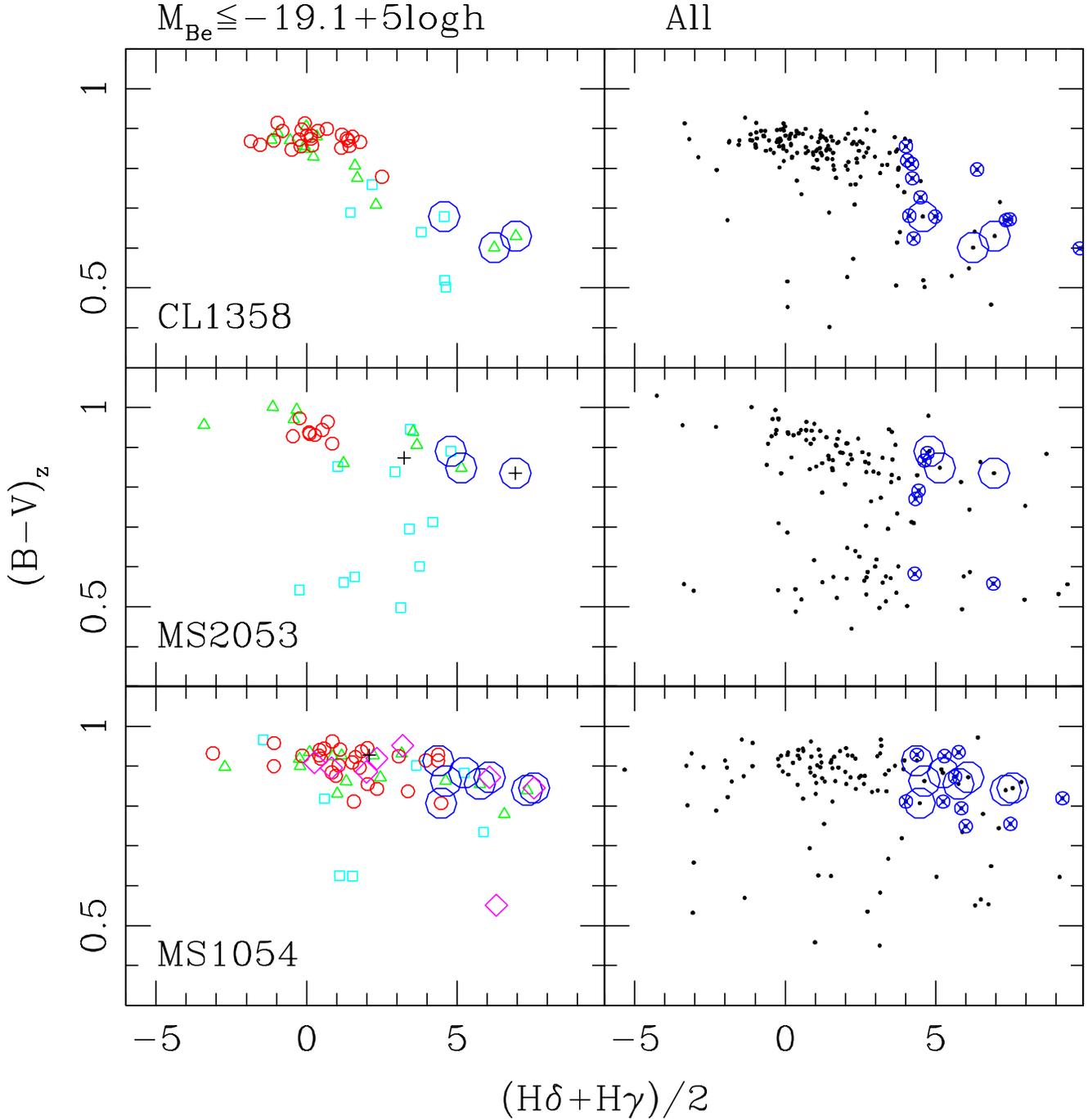}
\caption{{\it Left:} Comparison of $(B-V)_z$ versus Balmer absorption
[defined here as (H$\delta+$H$\gamma )/2$] for all members 
brighter than $M_{Be}=-19.1$\logh~in CL1358 ($z=0.33$; top),
MS2053 ($z=0.58$; middle), and MS1054 ($z=0.83$; bottom).  The
symbols represent different morphological types: E-S0 (small open
circles), S0/a-Sa (open triangles), spiral (open squares), no type (plus
signs), merger (diamonds), and E+A (large open circles).  E+A's tend
to be bluer than the early-type members.  {\it Right:} The same
comparison except now with all the confirmed cluster members.  Circles
with crosses denote E+A candidates fainter than our adopted magnitude
limit ($M_{Be}=-19.1$\logh) and/or with spectral $S/N<20$.
\label{BVz_balmer}}
\end{figure}

Figure~\ref{BVz_balmer} shows $(B-V)_z$ versus
$($H$\delta+$H$\gamma)/2$ for the cluster sample; morphological types
are included \citep{fabricant:00,fabricant:03}.  If we consider only
the galaxies brighter than our imposed magnitude limit, the average
color of E+A galaxies tends to be bluer than that of the early-type
population in each cluster.  As simple errors in the spectroscopy could
not produce such a uniformly blue sample in all three clusters, the E+A
fraction must be real and robust.


\subsection{The E+A Fraction in Intermediate Redshift Clusters}

For members brighter than our magnitude cut ($M_{Be}=-19.1$\logh), we
obtain E+A fractions of $7\pm4$\%, $10\pm6$\%, and 13$\pm5$\% at
$z=0.33,~0.58,~\&~0.83$ respectively (Table~\ref{eafractions}); note we
apply our strict E+A selection criteria here.  If we include all
confirmed members in each cluster and include all E+A candidates, the
fractions are $9\pm2$\%, $7\pm2$\%, and $16\pm3$\%.  Errors are
determined by assuming a Poisson distribution for the E+A galaxies.

Considering the low E+A fraction in Coma\footnote{Note that
\citet{caldwell:93} selected E+A's using different selection criteria
from a sample of predominantly early-type members.  However, recent
results from the WINGS survey \citep{poggianti:01,poggianti:03}
confirm the low E+A fraction in Coma for members brighter than
$M_v=-18.5$.}  \citep[$\lesssim3$\%;][]{caldwell:93}, these results
show the E+A fraction evolves strongly with redshift.  They also
suggest the E+A fraction continues to increase at $z>0.5$.  However,
larger samples at $z>0.3$ are needed to determine if this increase is
real or if the trend flattens at $z>0.3$.


\subsection{Comparison to CNOC1}

As noted earlier, there is disagreement in the literature as to whether
the E+A fraction in intermediate redshift clusters ($z\gtrsim0.3$) is
significantly higher than that of the field.  Published cluster surveys
($0.3<z<0.6$) estimate cluster E+A fractions of $\sim10-20$\%
\citep[D99]{belloni:95,couch:98}.  In contrast, B99 argue that the
average cluster E+A fraction is negligible ($1.5\pm0.8$\%) at
$z=0.18-0.55$, and comparable to the field.  Is it possible to
reconcile these two remarkably different claims?

While both D99 and B99 spectroscopically select their E+A samples using
H$\delta$ and [OII]~$\lambda3727$, B99 apply a correction that
decreases their cluster E+A fraction.  Recognizing that quiescent
galaxies with low Balmer indices dominate the cluster populations, B99
argue measurement errors will automatically produce outliers that are
classified as E+A's.  To remove this artificial inflation, they correct
the fraction of ``raw" cluster E+A's from $4.4\pm0.7$\% to
$1.5\pm0.8$\%.  Furthermore, B99 note that in CL1358 the E+A fraction
determined by F98 ($4.7\pm1.9$\%) is remarkably similar to the
uncorrected E+A fraction from their combined cluster sample
($4.4\pm0.7$\%).  They suggest F98 overestimated the true fraction by
not correcting for inclusion of spurious outliers.

\begin{figure}
\plotone{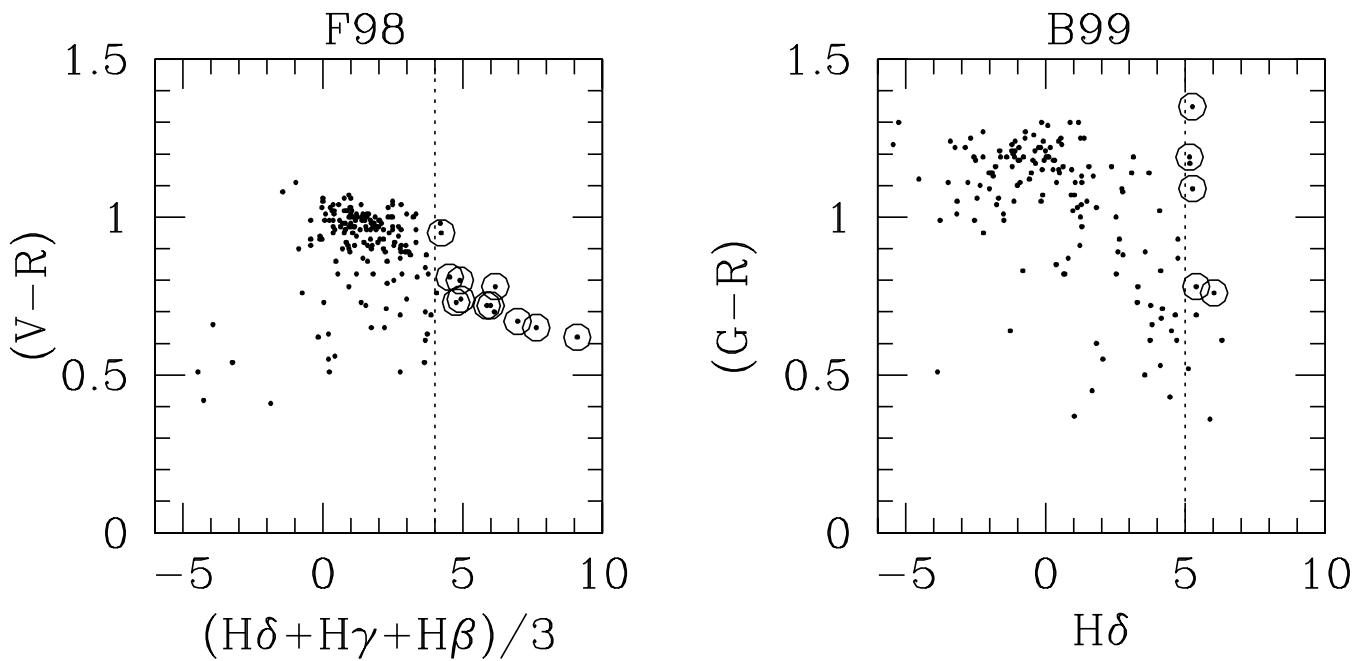}
\caption{Color versus Balmer strength for CL1358 members surveyed
  by F98 (left) and B99 (right); F98 and B99 used different
  ground-based filter systems.  Large open circles are E+A galaxies as
  defined in their respective surveys.  The two groups differ in how
  they define Balmer strength: F98 take the average of H$\delta$,
  H$\gamma$, and H$\beta$ while B99 only use H$\delta$
  (Table~\ref{ea_sel}).  The vertical dotted lines denote the Balmer
  cut-off for each survey.  With only one exception, E+A galaxies as
  defined by F98 are systematically {\it bluer} than the passive
  members.  In contrast, E+A galaxies selected by B99 span the range in
  color.  The E+A sample as determined by F98 is robust and
  demonstrates that E+A's make up a non-negligible fraction of cluster
  members ($\gtrsim5$\%).
\label{cnoc}}
\end{figure}

We test B99's theory by focusing on CL1358, a cluster included in B99's
analysis and spectroscopically surveyed by both CNOC1 \citep{yee:96}
and F98.  If a significant number of E+A's in this cluster are actually
passive galaxies, many of the E+A's should be as red as the
early-types.  Figure~\ref{cnoc} shows the distribution of color versus
Balmer strength for cluster members taken from F98 and B99; note B99
use only H$\delta$ as their Balmer criterion while F98 use the average
of H$\delta$, H$\gamma$, and H$\beta$ (Table~\ref{ea_sel}).  While
E+A's selected by B99 have a large spread in color and can be as red as
the passive population, virtually all of the E+A's found by F98 are
{\it bluer} than the passive early-types.  Simple errors in the
spectral indices could not produce such a uniformly blue sample, and so
these cannot be spectroscopically passive galaxies mistaken as E+A's.


In comparing results from F98 and B99 for $\sim100$ common members, we
find significant offsets in H$\delta$ and [OII] between the two
surveys; this may explain the discrepancy in the E+A fractions.  On
average, H$\delta$ equivalent width values are smaller ($\sim0.7$~\AA)
and [OII] emission larger ($\sim2$~\AA) in B99 as compared to F98.
Given the equivalent widths used to determine the E+A phase (see Table
\ref{ea_sel}), these offsets can seriously affect the final E+A
selection.  For example, if the indices of F98 are transformed to the
system of B99 and B99's selection criteria used, the fraction of E+A's
in CL1358 as measured with F98's spectra decreases from 4.7\% to 3.4\%.

B99 may have been motivated to correct their cluster E+A fractions if
large errors were associated with the CNOC1 H$\delta$ measurements.  To
determine if the CNOC1 errors are significantly larger than those of
F98 for CL1358, we independently estimate the equivalent width errors
of both samples by comparing them to the high signal-to-noise spectra
of K00b.  Using 27 members common to the three studies, we estimate the
true formal errors for H$\delta$ in both B99 and F98 are comparable
($\sim1.5$~\AA).  This value is much smaller than the H$\delta$ formal
error of 2.8\AA~associated with the F98 sample and subsequently used by
B99.  Thus in CL1358, B99 overestimated the correction factor due to
measurement errors.

To summarize, the discrepancy between B99 and F98 is due to 1) the
systematic offsets in the spectral indices, $i.e.$ B99's H$\delta$ EQW
values are smaller and [OII]$\lambda3727$ larger than F98's and 2)
B99's overestimate of the correction due to measurement errors.  As we
have shown, the offsets in the spectral indices used to define the
CNOC1 CL1358 E+A sample reduces the observed number of E+A's.  This
combined with B99's correction factor produces a very low cluster E+A
fraction.  The E+A fraction ($\sim5$\%) in CL1358 measured by F98 using
their stricter selection criteria is valid.

\section{Nature of Cluster E+A Galaxies}

Having established that a significant fraction of E+A galaxies exists
in clusters up to $z\sim0.8$, we now examine their physical properties
to determine what type of galaxies they are.  In the following, values
for quantitative structural parameters, $e.g.$ bulge-to-total fraction
($B/T$), galaxy asymmetry ($R_A$), total residual ($R_T$), half-light
radius ($r_{1/2}$), and bulge/disk scale length ($r_e, r_d$), were
measured by fitting two-dimensional de Vaucouleurs bulge plus
exponential disk surface brightness models to the galaxies
\citep{tran:02,tran:03a}.  The galaxy residuals $R_A$ and $R_T$ are
measured by by taking the difference between the HST images and
best-fit $r^{1/4}$ bulge+exponential disk model; how the residuals are
measured is explained more fully in \citet{tran:03a}.  Unless otherwise
noted, we only consider cluster members above our magnitude cut of
$M_{Be}=-19.1$\logh, and only the E+A's that satisfy our strict
selection criteria.

\subsection{Morphology}

In this sample, the cluster E+A's span the range in Hubble type
(Fig.~\ref{morph_hist}) and bulge-to-total fraction (Fig.~\ref{RA_bt})
to include both spirals and E/S0's.  However, the majority of E+A's
have measurable disks, consistent with results from previous cluster
E+A studies
\citep[D99]{wirth:94,couch:94,caldwell:97,couch:98,caldwell:99}.  Their
average bulge-to-total fraction of $\sim0.4$ reflects their tendency to
be disk-dominated systems.



This diverse range in $B/T$ and Hubble type is similar to the
heterogeneous morphologies found in studies of lower redshift cluster
E+A's \citep[$z\lesssim0.3$;][]{couch:98,caldwell:99}.  It also
confirms suggestions in earlier studies that E+A's must have a wide
variety of progenitors \citep[D99]{wirth:94,zabludoff:96}.  In
addition, the earliest-type E+A's are in our most distant cluster.  It
may be that more massive cluster members, i.e. early-types, had their
E+A phase at higher redshift.

\subsection{Interactions \& Mergers}\label{interactions}

The morphological variety of E+A galaxies suggests that they are
triggered by an external source, e.g. via the strong galaxy-galaxy
interactions or the tidal forces proposed in earlier studies
\citep{liu:95,belloni:95,zabludoff:96,caldwell:99}.  The cluster
environment provides a plethora of possible disruptive mechanisms, e.g.
ram-pressure stripping \citep{gunn:72}, gas compression
\citep{dressler:83}, perturbation by the cluster tidal field
\citep{byrd:90}, and galaxy harassment \citep{moore:96}.  Also, the
high merger fraction in MS1054 indicates galaxy-galaxy merging is
possible between members with low relative velocities
\citep{vandokkum:99}.  As these cluster E+A's are predominantly
disk-dominated systems, the more disruptive interactions would create
morphological signatures \citep[e.g.][]{barnes:92,moore:98} that are
visible during the E+A lifetime \citep[$\sim1.5$
Gyr;][]{couch:87,barger:96,leonardi:96}.  By examining the number of
E+A's that are considered mergers and/or that have high galaxy
residuals, we attempt to isolate which interactions, if any, are
associated with the E+A phase.

To identify mergers, we use classifications from \citet{vandokkum:99},
\citet{fabricant:00}, and \citet{fabricant:03}.  We consider high
residual galaxies as those having a high degree of asymmetry
\citep[$R_A\geq0.5$;][]{schade:95} and/or total residual
\citep[$R_T\geq0.1$;][]{tran:01}.

\begin{figure}
\plotone{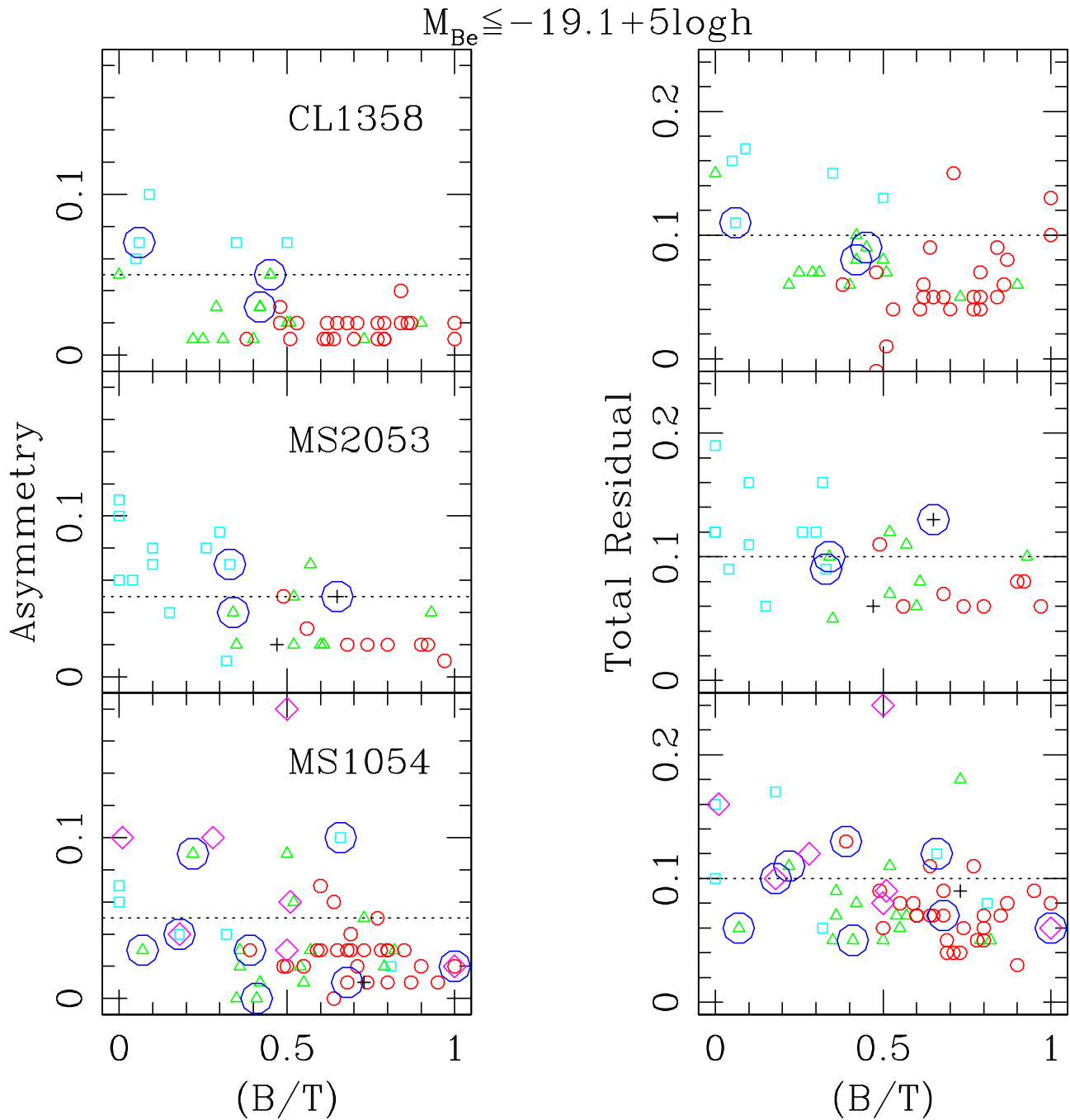}
\caption{Comparison of galaxy asymmetry ($R_A$; left panels) and total
  residual ($R_T$; right panels) to bulge fraction ($B/T$) determined
  from a de Vaucouleurs bulge+exponential disk profile.  Typical
  systematic errors for $R_A$, $R_T$, and $B/T$ are $\sim0.02,
  0.02,~\&~0.10$ respectively \citep{im:01,tran:02,tran:03a}.  Again,
  only galaxies above our magnitude cut are shown.  Approximately
  40--50\% are considered high residual galaxies, $i.e$ having either
  $R_A=0.05$ or $R_T=0.1$; these limits are denoted by dotted
  horizontal lines.  E+A's span the range in $B/T$ but the majority of
  them have $(B/T)<0.7$, meaning they have a measurable disk
  component.
\label{RA_bt}}
\end{figure}

Despite MS1054's high merger fraction
\citep[$\sim17$\%;][]{vandokkum:99}, only two of the mergers are
considered E+A's; no other E+A's in our sample are associated with
mergers.  \citet{wirth:94} and D99 also observe a low incidence of
mergers associated with cluster E+A's.  We find only about half of the
cluster E+A's have high galaxy residuals (Fig.~\ref{RA_bt}).  For
comparison, the fraction of high residual E+A's is larger than that of
early-types (E-S0; $<15$\%) and even early-type spirals (S0/a-Sa;
$\sim30$\%) but less than that of cluster spirals ($\sim80$\%).  The
number of cluster E+A's with prominent disks combined with only half
the sample having high galaxy residuals suggests mergers are not the
primary trigger of the E+A phase in cluster galaxies.

\subsection{Color-Magnitude Diagram}\label{cmd}

Even though galaxies can be brightened significantly during the E+A
phase \citep[up to $\sim1.5$ mag;][]{barger:96}, E+A's in nearby
clusters tend to be faint systems
\citep[$L\lesssim0.4L^{\ast}$;][]{caldwell:99}.  However, past studies
of intermediate redshift clusters find E+A's with $L> L^{\ast}$
\citep[D99]{wirth:94}.  Here we determine if the cluster E+A's in this
sample are as luminous as the brightest cluster members, and whether
they also cover a wide luminosity range.


In Fig.~\ref{BVz_MBz}, we show the color-magnitude (CM) distribution
of all cluster members and E+A candidates; all cluster members,
including the E+A candidates, have been corrected for simple passive
evolution (\S2.1).  In all three clusters, we find bright E+A's
($M_{Be}\lesssim-19.1$\logh); half of the 14 robust E+A's are brighter
than $M_{Be}^{\ast}$ \citep[$-19.5$\logh~at $z=0.83$;][]{hoekstra:00}.
Even more striking are the very luminous E+A's at $z=0.83$: these
E+A's are up to a magnitude brighter than their lower redshift
counterparts and cover a larger magnitude range.  Note the cluster
E+A's tend to be bluer than the red sequence.  The E+A luminosity
range, particularly at $z=0.83$, only reinforces the conclusion that
they have a heterogeneous parent population.  The fact that the
brightest E+A's in this sample are in our most distant cluster is
additional evidence for down-sizing of the cluster E+A population.

\subsection{Brightening During the E+A Phase}\label{bright}

Simple models show that during the post-starburst phase, galaxies can
be brightened up to 1.5 magnitudes in the optical
\citep{newberry:90,barger:96}.  For comparison, we place here
observational constraints on $\Delta M_{Be}$ using the internal
velocity dispersions ($\sigma$) acquired for 120 members and the
Fundamental Plane \citep[see Appendix]{djorgovski:87,faber:87}. As
demonstrated in e.g. \citet{faber:87}, residuals from the FP can be
expressed as residuals in the $M/L$ ratio.  We find the
$\log(M/L)$ residuals of the nine cluster E+A's with measured $\sigma$
range from $\sim-0.5$ to $\sim0.3$ (Fig.~\ref{dlgML_dBV}).  Assuming
E+A's fade until $\Delta\log(M/L)=0$, we estimate cluster E+A's are
brightened by as much as $\Delta M_{Be}\sim1.25$ mag, with a median of
$0.25$ mag.

\begin{figure}
\plotone{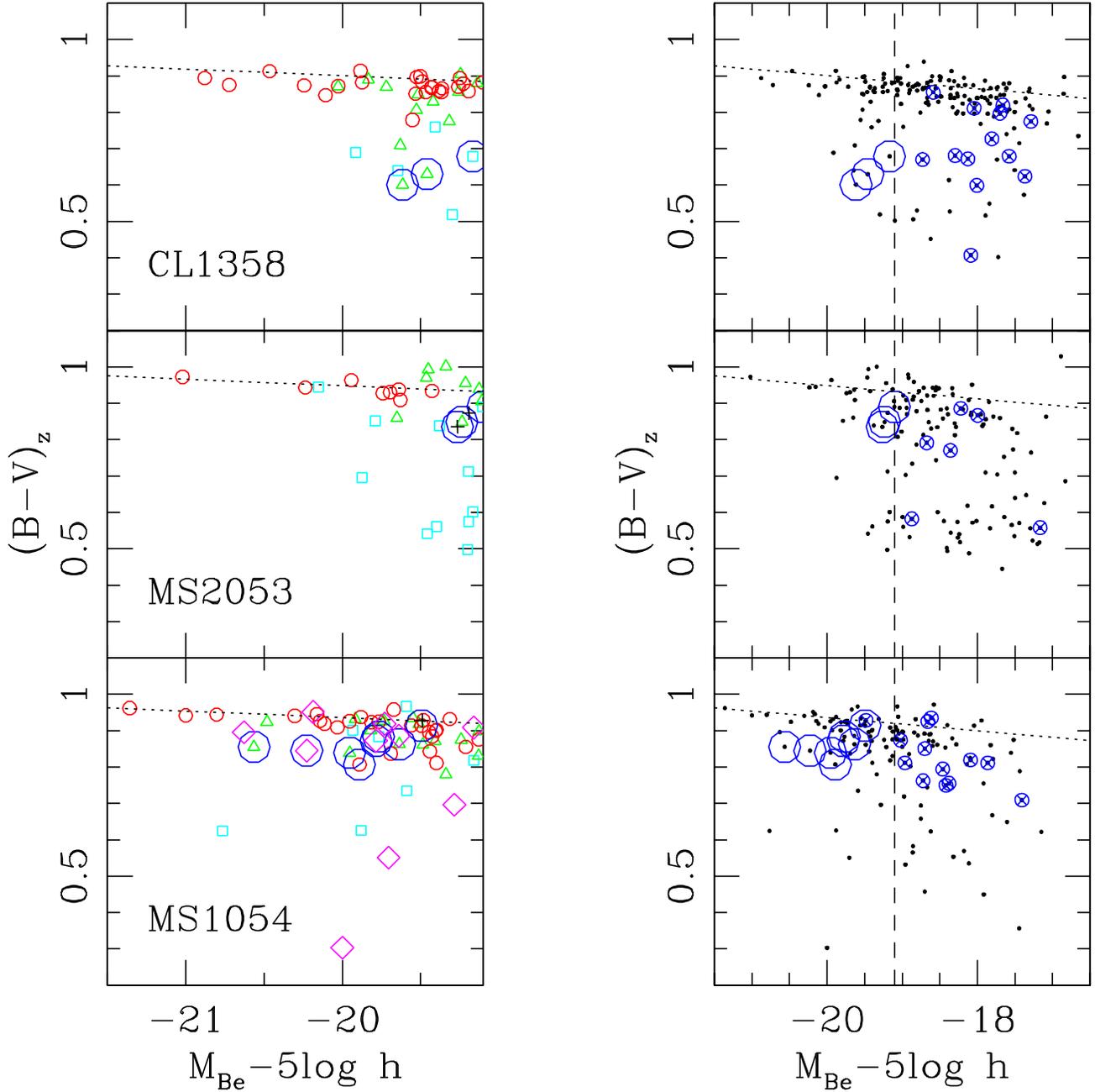}
\caption{The left panels show the color-magnitude diagram for members
  with $M_{Be}\leq-19.1$\logh~while the right panels show all confirmed
  cluster members (symbols are as in Fig.~\ref{BVz_balmer}).  The
  color-magnitude relation (dotted line) is normalized to the E-S0
  members and its slope is from \citet{vandokkum:00}.  E+A galaxies in
  the two lower redshift clusters are faint ($M_{Be}\gtrsim-19.5$\logh)
  but can be up to a magnitude brighter in MS1054.
\label{BVz_MBz}}
\end{figure}

Assuming the E+A's at $z=0.83$ redden and fade by $\sim0.25$ mag by
$z=0.33$, the only galaxies in CL1358 in this luminosity and color
range are E-S0's and S0/a-Sa's (Fig.~\ref{BVz_MBz}).  This suggests
that some of the brightest early-type galaxies in nearby clusters had
an E+A phase in their past.


\subsection{Internal Velocity Dispersions}

Having demonstrated that E+A's at higher redshift can be as luminous
as the brightest cluster members (\S\ref{cmd}), we now determine if
these brighter E+A's are also {\it massive} galaxies, or whether they
are simply low luminosity/mass members that are temporarily
brightened.  By determining the E+A mass distribution, we can
characterize what the E+A progenitors are and also constrain what
their descendants at lower redshift can be.  To address this issue,
diagnostics that are not likely to depend strongly on redshift are
needed, $i.e.$ internal velocity dispersions and half-light radii.

While half-light radii ($\sim$ sizes) are measured fairly robustly from
the WFPC2 imaging \citep{tran:03a}, determining internal velocity
dispersions ($\sigma$) at these redshifts is challenging.  However, we
have obtained direct $\sigma$ measurements for 120 cluster members
\citep[K00b]{kelson:97,vandokkum:98b,kelson:01}.  With the colors,
effective radii\footnote{To estimate velocity dispersions, we fit a
  pure de Vaucouleurs profile to the members instead of our normal de
  Vaucouleurs bulge+exponential disk.}, luminosities, and measured
dispersions, we estimate velocity dispersions for the rest of the
cluster sample using the Fundamental Plane.  In this method, we correct
$M/L$ ratios of later-type members and essentially evolve them onto the
color-magnitude relation defined by the early-types; see the Appendix
for a detailed explanation of this method.

\begin{figure}
\plotone{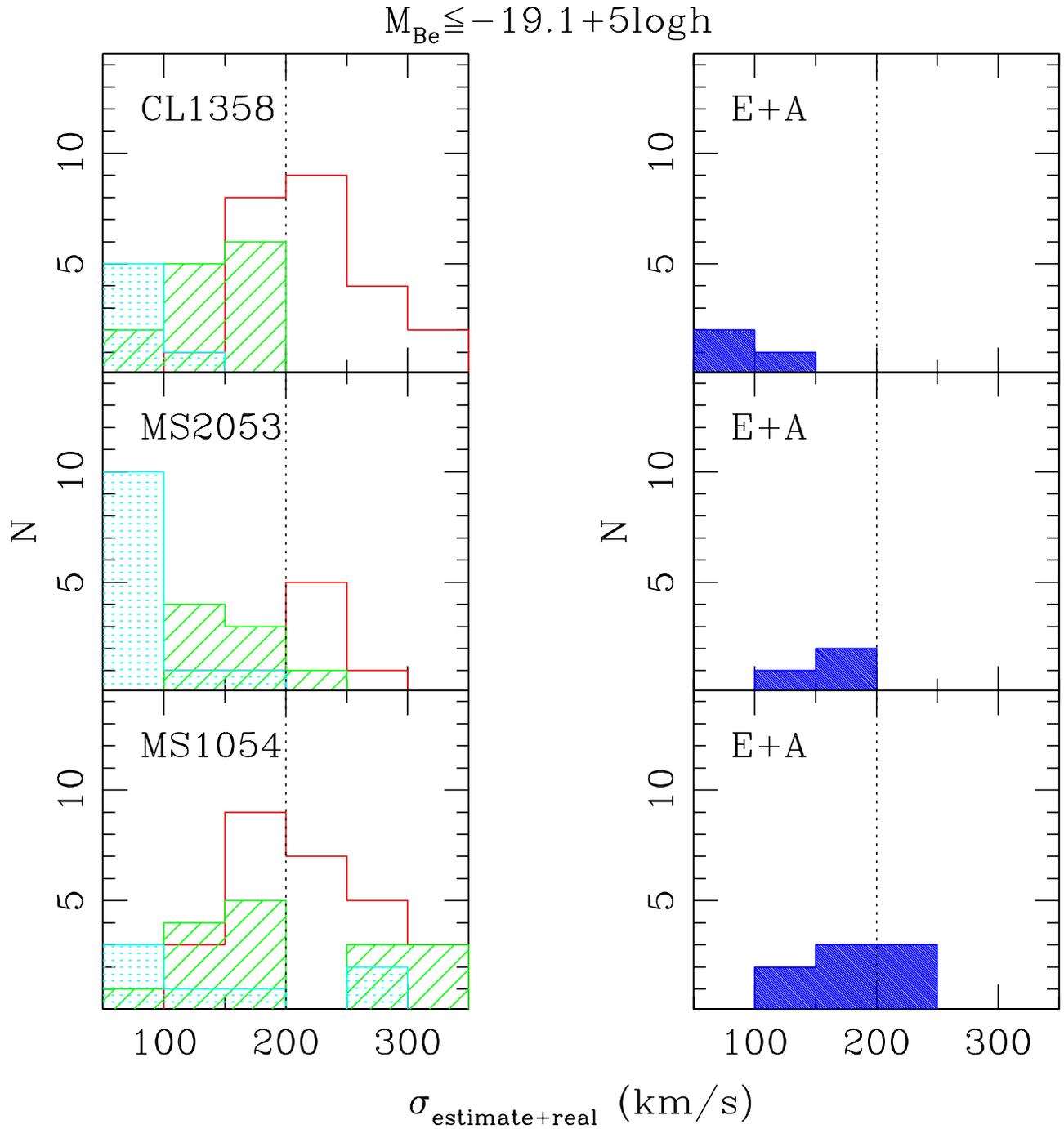}
\caption{Histogram of internal velocity dispersions (measured and
  estimated) for members brighter than our magnitude limit.  In the left
  panels, open regions correspond to E-S0's ($-5\leq T\leq-1$), hatched
  regions to S0/a-Sa's ($0\leq T\leq1$), and dotted regions to spirals
  ($T\geq2$).  The right panels show the respective E+A's that satisfy
  our strict selection criteria.  All of the high $\sigma$ ($>200$\kms;
  dotted vertical line) members in CL1358 are E-S0's.  In contrast,
  high $\sigma$ members in MS1054 include S0/a-Sa's, E+A's, spirals, and
  mergers (not shown).
\label{nsigma_hist}}
\end{figure}

Figure~\ref{nsigma_hist} shows the distribution of internal velocity
dispersions (measured and estimated $\sigma$) for cluster members
brighter than our magnitude cut.  The range in velocity dispersion for
E+A galaxies increases at higher redshift: E+A's at $z=0.33$ have
smaller velocity dispersions ($\sigma\lesssim 150$\kms) than at
$z=0.58$ ($\sigma\lesssim 200$\kms) and $z=0.83$ ($\sigma\lesssim
250$\kms).  The difference between $z=0.33$ and $z=0.83$ is most
striking.  Considering the robustness of the spectroscopic data and
high quality of the WFPC2 imaging, any E+A's with $\sigma>200$\kms
(measured or estimated) in the two lower redshift clusters would have
been easily detected {\it if they existed}.

\begin{figure}
\plotone{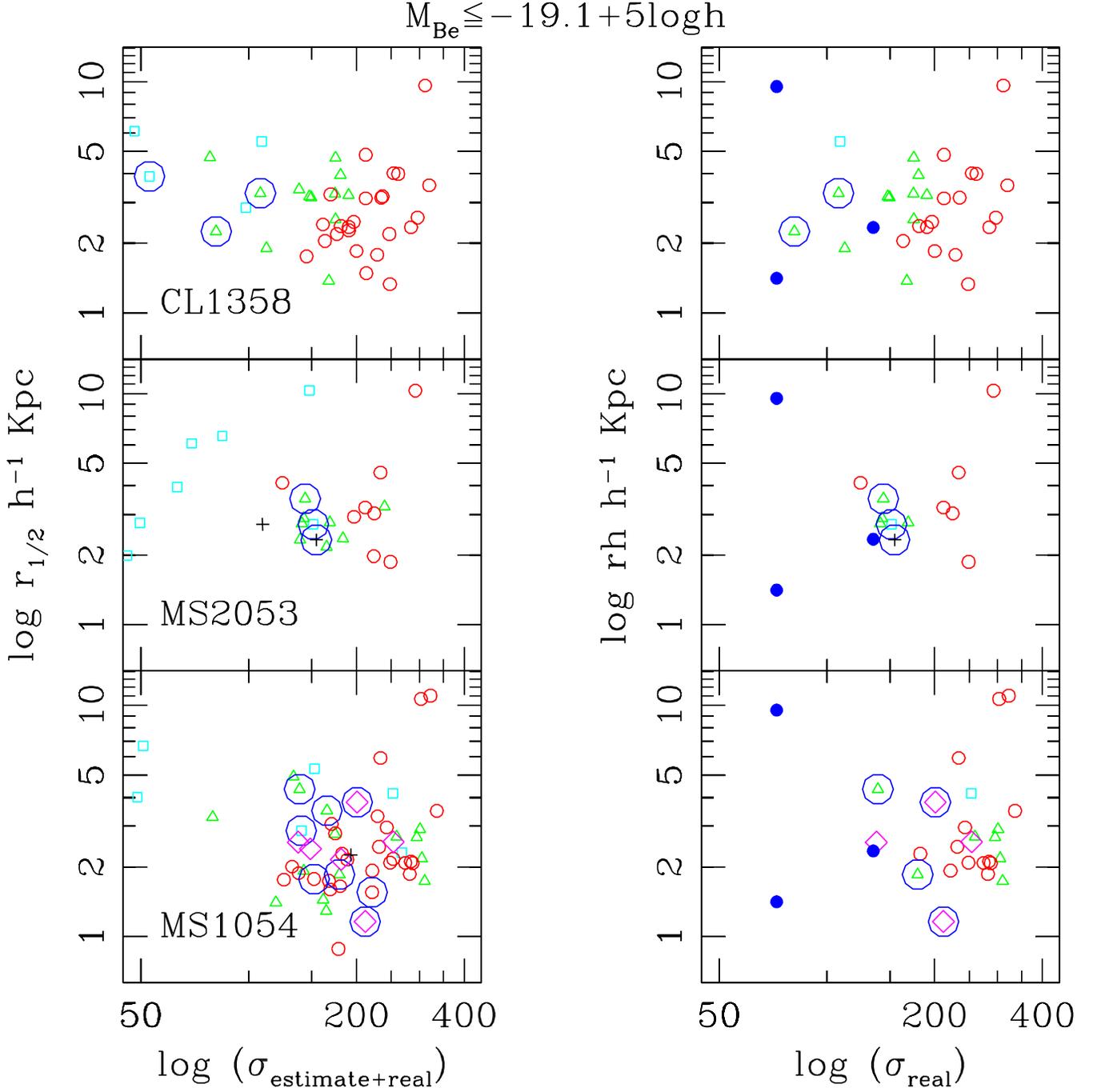}
\caption{Galaxy size, as defined by half-light radius from a de
  Vaucouleurs bulge+exponential disk model, compared to velocity
  dispersion for $M_{Be}\leq-19.1$\logh~members; both these quantities
  should be relatively constant with redshift.  The symbols are as in
  Fig.~\ref{BVz_balmer}.  The left panels utilize estimated velocity
  dispersions (except when actually measured) while only members with
  measured $\sigma$ are shown in the right panels.  Included in the
  right panels (solid dots) for comparison are E+A's in Coma; here we
  use velocity dispersions from \citet{caldwell:96} and half-light
  radii from \citet{scodeggio:98}.  The high $\sigma$ E+A's in MS1054
  are very likely to be the progenitors of massive early-types at low
  redshifts.  Also note the number of high dispersion
  ($\sigma>250$\kms) S0/a-Sa's (triangles) in MS1054; the only possible
  counterparts to these systems in CL1358 are E-S0's.
\label{rh_lognsigma}}
\end{figure}

To emphasize the disparity between the three clusters, we compare their
half-light radii to internal velocity dispersions in
Fig.~\ref{rh_lognsigma}; neither of these parameters is expected to
evolve significantly within the redshift range covered here.  For
comparison, we also include E+A's from Coma using velocity dispersions
from \citet{caldwell:96} and half-light radii from
\citet{scodeggio:98}.  The half-light radii and dispersions of E+A's at
$z=0.83$ are comparable to those of the massive early-type members in
all three clusters.  In contrast, none of the E+A's at $z=0.33$ nor in
Coma could be considered a cluster giant.  It is apparent that the
progenitors of the lower redshift E+A's are very different from those
at $z=0.83$.  As with their luminosity, the E+A's with the highest
internal velocity dispersions are found in our most distant cluster.

Another key result from our analysis is that in the lowest redshift
cluster, we find no counterparts to the bright, high $\sigma$
late-types found at $z=0.83$.  The $\sigma>200$\kms~galaxies in MS1054
include S0/a-Sa's, E+A's, spirals, and mergers while the only galaxies in
CL1358 with such high dispersions are E-S0's (Figs.~\ref{nsigma_hist}
\& \ref{rh_lognsigma}).  If we assume that, given their similar cluster
dispersions (Table~\ref{clusters}), CL1358 ($z=0.33$) is an evolved
version of MS1054 ($z=0.83$), then the wide mix of high $\sigma$
systems in MS1054 must be morphologically transformed into E-S0's
within $\sim2.5$ Gyr.  The E+A phase may be an integral step in this
process.

\subsection{Spatial Distribution \& Substructure}

\begin{figure}
\epsscale{1}
\plotone{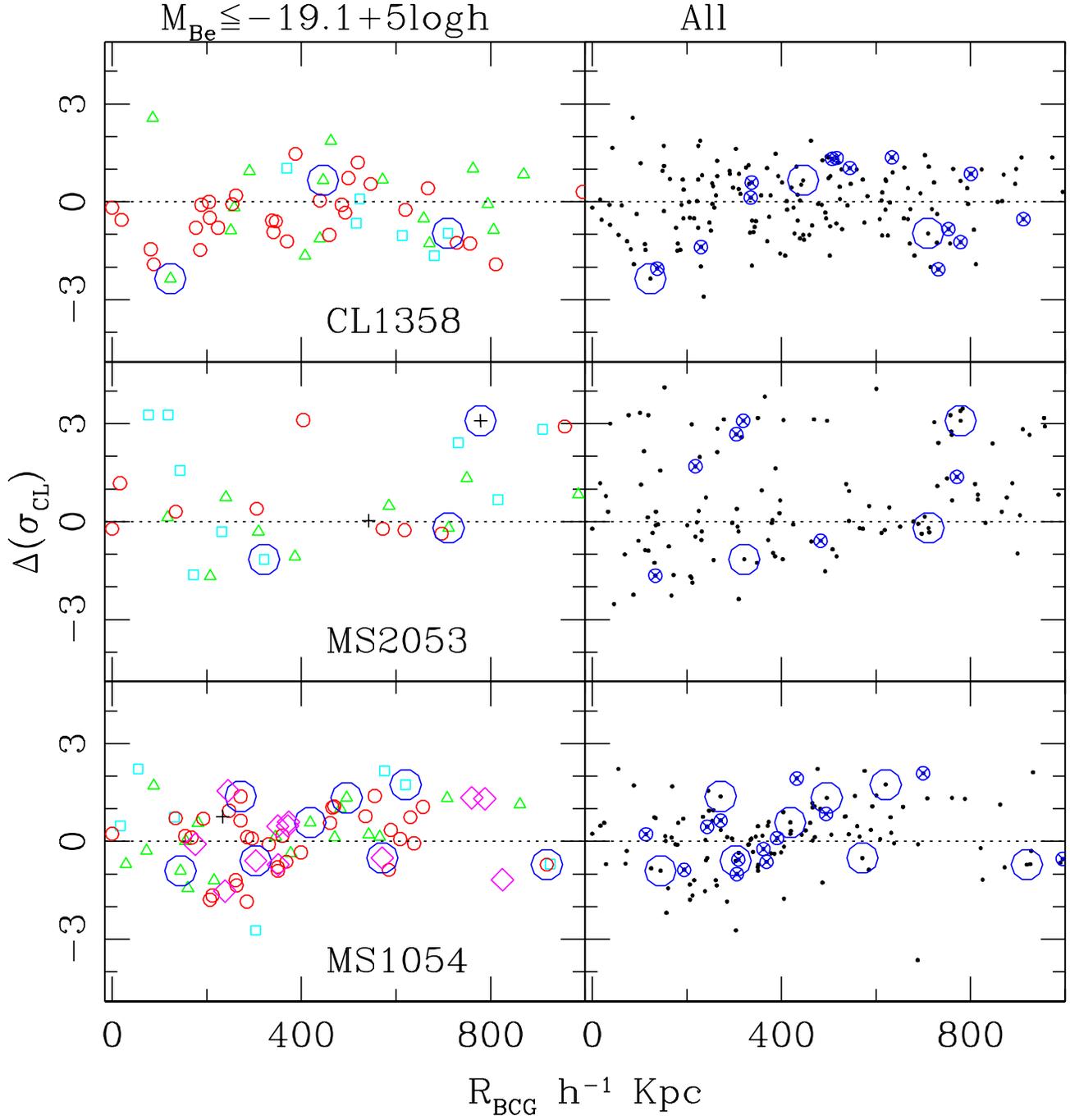}
\caption{Deviation from the cluster's mean velocity dispersion in
  units of $\sigma_{CL}$ (the cluster velocity dispersion) for each
  member.  The left panels show only members with
  $M_{Be}\leq-19.1$\logh~while the right panels show all confirmed
  members; the symbols are as in Fig.~\ref{BVz_balmer}.  The large
  subcluster in MS2053 \citep{tran:02} is evident in the middle right
  panel.  In this sample, E+A's are distributed throughout the cluster
  but none are found at $R_{BCG}<100$\hi kpc.
\label{dzsigma_Rbcg}}
\end{figure}

The E+A's in our sample are found at $R_{BCG}$ of $\sim100-900$\hi kpc
(Fig.~\ref{dzsigma_Rbcg}).  Like D99, we find that E+A's tend to avoid
the inner cluster core ($R_{BCG}\lesssim100$\hi kpc).  At $z=0.83$,
three of the eight E+A's are associated with a large subcluster
\citep[$>20$ members;][]{tran:02}, while at $z=0.58$ the E+A's are
found in both the main cluster and massive subcluster
(Fig.~\ref{dzsigma_Rbcg}, middle right).  This suggests that
$\sim30$\% of E+A's, if not more, are associated with the groups that
are being accreted by the clusters.  



\section{Discussion}

In the following section, we attempt to form a coherent picture of the
cluster E+A population at intermediate redshifts.  Using the physical
properties detailed in \S4, we determine if E+A's would be equally as
numerous in a mass selected sample and test if E+A's are drawn from
the same parent population as regular cluster members\footnote{We
consider the regular cluster population to follow the luminosity
functions determined from weak-lensing studies by
\citet{hoekstra:98,hoekstra:00,hoekstra:02}.  Here members were
photometrically selected to have color deviations of $<0.2$ mags from
the color-magnitude relation normalized to early-type members.}, and
establish a connection between the progenitors and descendants of
these systems.

As in \S4, we only consider cluster members brighter than
$M_{Be}=-19.1$\logh, and E+A's that satisfy our strict selection
criteria (\S3.2).  One possible concern is how the varying richness of
the three clusters affects the conclusions drawn from these data.
However, we emphasize it is the relative number of spectra that is
important.  Our large sample of confirmed cluster members
($>120$/cluster) combined with the extensive spectroscopic and
photometric properties we have gathered allows us to make a meaningful
analysis of the cluster E+A population.

\subsection{Luminosity vs. Mass Selected E+A Sample}

We find the fraction of E+A galaxies in intermediate redshift clusters
ranges from $7-13$\% (Table~\ref{eafractions}).  However, we note our
spectroscopic survey is magnitude limited.  From \S\ref{bright}, we
know E+A's can be brightened by as much as $\Delta M_{Be}\sim1.25$ mag,
and so the E+A fraction in a {\it mass selected} cluster sample might
be lower.  Here we determine the influence of brightening on the E+A
fraction and estimate a mass selected fraction.

\begin{figure}
\epsscale{0.5}
\plotone{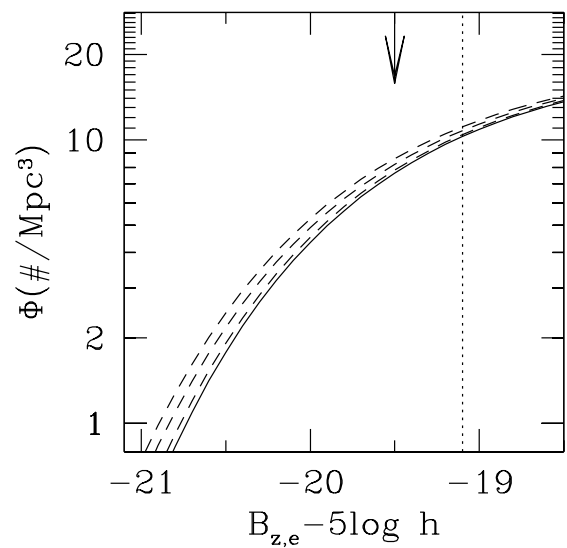}
\caption{How the total luminosity function (solid line) varies when
  10, 30, and 50\% of the members are brightened by $M_{Be}=0.25$ mag
  (dashed lines, increasing from the right); the $\Phi$ normalization
  is arbitrary.  The total luminosity function is a combination of a
  regular distribution ($M_{Be}^{\ast}=-19.5$\logh; arrow) and a
  brightened distribution ($M_{Be}^{\ast}=-19.75$\logh).  The dotted
  vertical line shows our imposed magnitude limit
  ($M_{Be}=-19.1$\logh).  Even if only 10\% of the members are
  brightened, the E+A fraction of a luminosity selected sample is $\sim30$\%
  larger than that of a mass selected one.
\label{lumfunc}}
\end{figure}

We first use the Schechter luminosity function \citep{schechter:76} to
populate cluster members as a function of magnitude; at $z=0.83$,
$M_{Be}^{\ast}=-19.5$\logh~mag and $\alpha=-1$ \citep{hoekstra:00}.
Since E+A's are brightened by $<\Delta M_{Be}>_{med}=0.25$ mag (see
\S\ref{bright}), they follow a luminosity function with
$M_{Be}^{\ast}=-19.75$ mag.  By combining the two luminosity
functions, we can estimate approximately how biased a luminosity
selected sample is (Fig.~\ref{lumfunc}).  Note this approach assumes
1) all E+A's are brightened by $0.25$ mags and 2) E+A's have the same
$\alpha$ as regular cluster members.  If 10\% of the members are
E+A's, i.e.  if the total cluster luminosity function comprises 90\%
regular and 10\% brightened, we estimate the E+A fraction in a
luminosity selected sample ($M_{Be}\leq-19.1$\logh) is $\sim1/3$
larger than that of a mass selected sample.  Depending on the
magnitude limit, the E+A fraction in a mass selected sample can differ
by $\sim30$\% compared to a magnitude selected fraction.

\subsection{Progenitors and Descendants}\label{descendants}

In Coma, known E+A's are low luminosity ($L<0.4L^{\ast}$), low
dispersion ($\sigma<150$\kms) systems that are unlikely to evolve into
massive early-type members
\citep{caldwell:96,caldwell:99,poggianti:03}.  However, we find this
is not the case at $z>0.3$.  From Figs.~\ref{nsigma_hist} \&
\ref{rh_lognsigma}, we see that at $z=0.33$, E-S0's and S0/a-Sa's are
the only logical descendants of the high dispersion ($\sigma>150$\kms)
E+A's at $z=0.83$.  The E+A phase may signify the transformation of
early-type spirals into E-S0's, and strongly star-forming spirals into
S0/a-Sa's.

The young stellar ages implied by the E+A phase may seem to conflict
with the old stellar ages ($z_f>2$) derived from studies using the FP
and absorption line strengths
\citep{kelson:97,vandokkum:98b,kelson:01}.  However, these ages
represent the mean epoch of star formation and do not preclude activity
at $z<2$.  Furthermore, it is not clear whether all galaxies undergo
the E+A phase.  Also note that the total starburst population can be as
little as 10\% of the galaxy's final stellar mass \citep{barger:96}.
Assuming the majority of their stars formed at $z_f>2$, even early-type
members can be E+A's.

The connection between E+A progenitors and descendants agrees very well
with the concept of ``progenitor bias'' introduced by
\citet{vandokkum:01}.  In this scenario, as many as 50\% of present day
early-type members are transformed from (later) galaxy types at $z<1$.
This morphological evolution is strongly supported by the likely
transformation of the E+A's at $z=0.83$ to early-type members by
$z=0.33$.

\begin{figure}
\epsscale{1}
\plotone{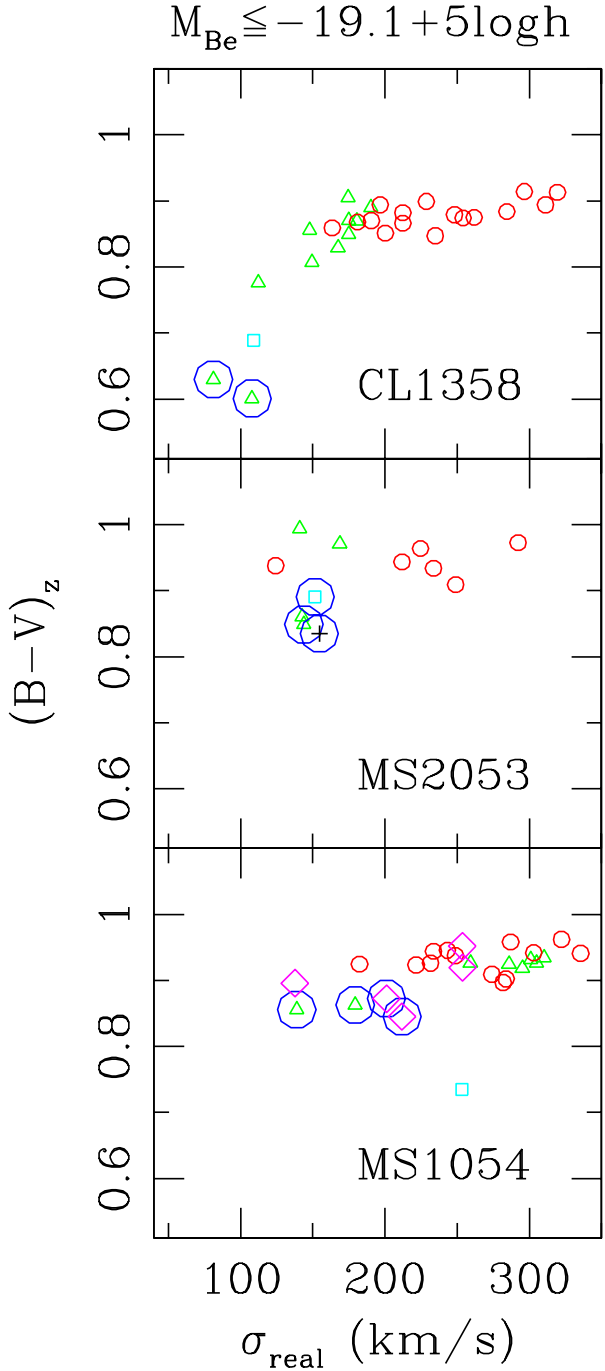}
\caption{Color versus internal velocity dispersion for only cluster
  members with {\it measured} dispersions that are brighter than our
  magnitude limit ($M_{Be}=-19.1$\logh).  At $z=0.33$, the only
  counterparts to the passive, high dispersion ($\sigma>250$\kms)
  S0/a-Sa's (triangles) in MS1054 are E-S0's (small circles).  Also note
  the logical descendants of $\sigma\sim200$\kms~E+A's in MS1054 are
  early-types.  Both these points are compelling evidence for
  morphological evolution in cluster members.
\label{BVz_sigma}}
\end{figure}

Additional evidence for morphological transformation is provided by the
high dispersion ($\sigma>250$\kms) S0/a-Sa members at $z=0.83$
(Fig.~\ref{BVz_sigma}).  The {\it only} $z=0.33$ counterparts to these
systems are E-S0 members.  Although red early-type spirals have been
observed in lower redshift clusters, e.g.  \citet{bower:92} and
K00b, none have dispersions as high as those in MS1054.

\subsection{E+A ``Down-sizing''}

In our cluster E+A sample, we find a trend of decreasing luminosity
and internal velocity dispersion with decreasing redshift.  This
evolution is similar to the observed decrease since $z\sim1$ in the
maximum luminosity of field galaxies undergoing rapid star formation
\citep[``down-sizing'';][]{cowie:96}.  While a similar trend has also
been suggested for clusters
\citep{poggianti:01,kodama:01,poggianti:03}, here we find compelling
evidence that down-sizing applies directly to cluster E+A galaxies.
The E+A's in our most distant cluster ($z=0.83$) can be luminous ($L$
up to $2.5L^{\ast}$) and can have high internal velocity dispersions
($\sigma>200$\kms).  In contrast, the E+A's in our lowest redshift
cluster ($z=0.33$) are $L\lesssim L^{\ast}$ and have low dispersions
($\sigma<150$\kms).  The data are such that if any bright, high
dispersion E+A's existed in our lower redshift clusters, we would have
detected them.

One possible concern is that we may be observing an evolution in the
E+A {\it number fraction} rather than a real evolution in mass (as
traced by $\propto r_e\sigma^2$).  If the E+A fraction increases with
redshift, the likelihood of observing a luminous, high dispersion E+A
also increases.  With only three clusters, we cannot determine with
certainty if the cluster E+A fraction continues to increase at $z>0.3$,
but we do note the E+A fraction is highest ($\sim13$\%) in the $z=0.83$
cluster.

\subsubsection{Statistical Tests}

We attempt to break the degeneracy between evolution in number fraction
and in mass by comparing the E+A magnitude, velocity dispersion, and
mass distributions to those of the spectroscopically confirmed cluster
population.  If E+A's are distributed like these other cluster members,
we expect to find no difference in the E+A luminosity, velocity
dispersion, and mass distributions when compared to the rest of the
cluster members.  Even if the E+A fraction evolves, this should be true
at all redshifts.  However, if we can establish that E+A's are
distributed differently from the other cluster members, this
would support evolution in mass.

\begin{figure}
\epsscale{1}
\plotone{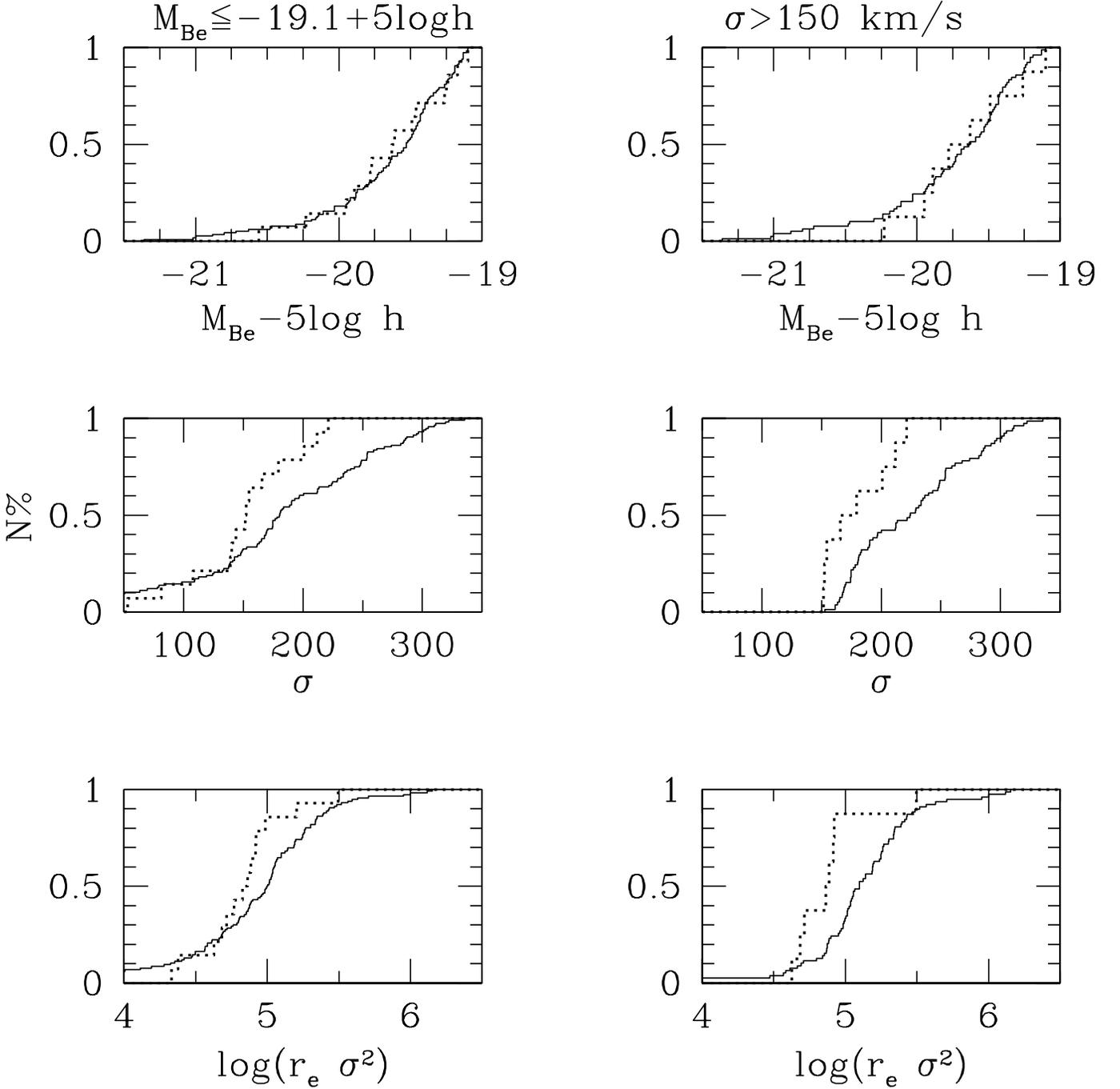}
\caption{{\it Left Panels:} To determine if the E+A population (dotted
  lines) differs from the cluster population (solid lines) as a whole,
  we compare their magnitude (top), internal velocity dispersion
  (middle), and mass (as traced by $r_e\sigma^2$; bottom) K-S
  distributions.  All three cluster samples are combined to improve the
  statistics, and we consider only members above our magnitude limit
  ($M_{Be}=-19.1$\logh).  We find the E+A and cluster $M_{Be}$
  distributions are indistinguishable using K-S and Wilcoxon tests.
  However, the same tests find the $\sigma$ and mass distributions are
  different at $\sim90$\% CL.  Compared to the rest of the
  spectroscopically confirmed cluster sample, these E+A's tend to have
  inherently lower velocity dispersions and masses.  {\it Right
    Panels:} The same analysis except now we apply a velocity
  dispersion limit ($\sigma>150$\kms) in addition to our magnitude cut.
  Again, we find E+A and cluster $M_{Be}$ distributions to be
  indistinguishable but their $\sigma$ and mass distributions differ
  (both at $>95$\% CL).
\label{KS_test}}
\end{figure}

In Fig.~\ref{KS_test}, we compare the luminosity, internal velocity
dispersion, and mass ($\propto r_e\sigma^2$) distribution of
spectroscopically confirmed cluster members to the E+A's
(Fig.~\ref{KS_test}, left panels).  Here we combine all three clusters
to improve the statistics and use the non-parametric Kolmogorov-Smirnov
\citep{press:92} and Wilcoxon tests \citep{siegel:88}, and consider
only members brighter than $M_{Be}=-19.1$\logh.  Both tests find the
E+A magnitude distribution to be indistinguishable from that of the
members.  Conversely, the same tests find the $\sigma$ and mass
distributions to be different with $\sim90$\% confidence.  Although
they can be as bright as the dominant early-types, the E+A's in this
sample tend to have lower velocity dispersions and masses when compared
to the rest of the cluster population.

Given we know galaxies are brightened during the E+A phase by as much
as $\Delta M_{Be}\sim1.25$ mag (\S\ref{bright}), we note cluster E+A's would
automatically be expected to have lower dispersions in a magnitude
limited sample.  To remove this possible bias, we now apply a velocity
dispersion limit of $\sigma>150$\kms~in addition to our magnitude limit
(Fig.~\ref{KS_test}, right panels).  Again, we find the luminosity
distributions of E+A's and cluster members to be indistinguishable
while their dispersion and mass distributions are different ($>95$\% CL
for both).  These results indicate the increase in cluster E+A
luminosity and velocity dispersion with redshift reflects evolution in
the E+A mass distribution rather than evolution in their number
fraction.

\begin{figure}
\epsscale{0.6}
\plotone{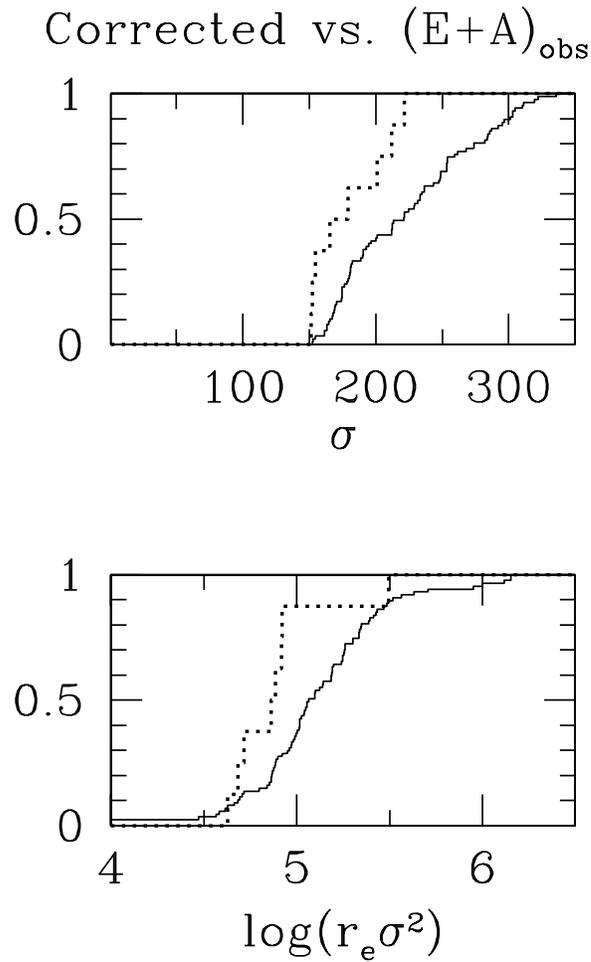}
\caption{We perform a modified K-S test on the observed E+A
  sample (dotted lines) to the cluster sample (solid lines) corrected
  for incompleteness ($M_{Be}\leq-19.1$\logh).  We use the combined
  cluster sample to improve our statistics and consider only galaxies
  with $\sigma>150$\kms.  The difference between the E+A's and the
  other cluster members is significant at the 98\% CL for both the
  internal velocity dispersion and mass ($\propto r_e\sigma^2$)
  distributions.  The significance was derived from Monte Carlo
  simulations (5000 realizations).
\label{ksfull}}
\end{figure}

To remove any possible bias introduced by spectroscopic
incompleteness\footnote{By $M_{Be}=-19.1$\logh, the completeness of the
  spectroscopic surveys drops to $\sim60-70$\%.}, we now estimate a
complete cluster sample by weighting each member by the inverse of the
cluster's magnitude selection function \citep{vandokkum:00,tran:02}.
Here we assume the worst case scenario where the cluster sample is
incomplete but the E+A sample is complete; in this case, we would
expect the minimum difference between the two populations.  As before,
we consider only galaxies with $M_{Be}\leq-19.1$\logh~and
$\sigma>150$\kms.  We find again a large difference between the two
distributions of the velocity dispersions and masses
(Fig.~\ref{ksfull}).


We establish the significance of the difference by deriving the
probability distribution of the $D_{max}$ statistic in this modified
K-S test from simulations.  We assume that the parent populations are
the same, but that the spectroscopically confirmed cluster sample
suffers from incompleteness for which the counts are corrected.  From
5000 realizations, we find that the probability of finding $D_{max} >
D_{max}(observed)$ is lower than 2\% for both the velocity dispersion
and mass distributions.  Thus, it is very unlikely that the E+A's are
drawn from the same parent population as the rest of the cluster
members.

\subsubsection{Independent Observational Evidence}

\begin{figure}
\epsscale{0.6}
\plotone{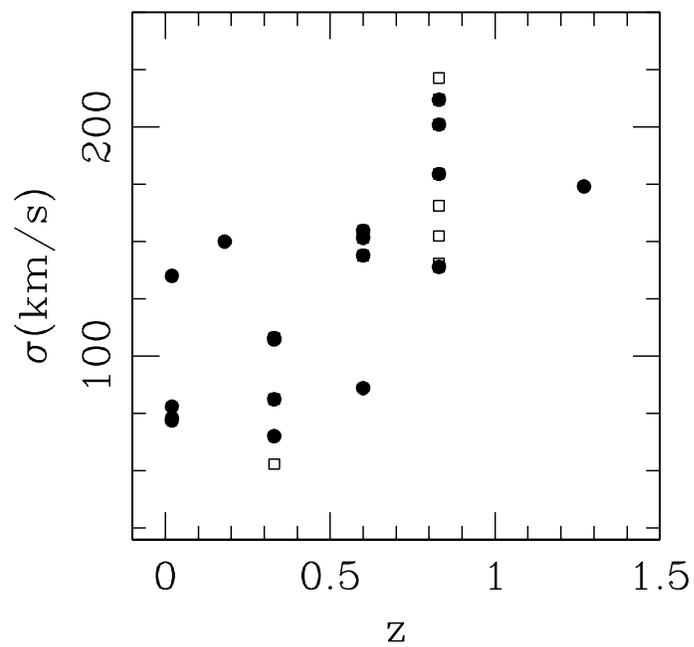}
\caption{Distribution of internal velocity dispersions for cluster
  E+A's from our sample and from the literature
  \citep[$z=0.02,~0.18,~1.27$;][]{caldwell:96,franx:93,vandokkum:03}.
  Filled and open circles represent measured and estimated velocity
  dispersions respectively.  The trend of increasing dispersion with
  redshift strongly suggests the mass distribution of cluster E+A's
  evolves with redshift (``down-sizing'').
\label{sigma_z}}
\end{figure}

We find additional evidence of mass down-sizing by combining our E+A
sample with results from the literature (Fig.~\ref{sigma_z}).  While
high $\sigma$ E+A's are found in clusters up to $z\sim1.3$
\citep{vandokkum:03}, it seems only low $\sigma$ ones exist in nearby
clusters, e.g. Coma.  We also find a trend of increasing $\sigma$ with
redshift in the range $0.02<z<1.27$.  This increase in the
characteristic E+A mass with redshift is consistent with the trend of
increasing age with increasing velocity dispersion found in Coma
early-types by \citet{caldwell:03}.  

Note these results do not exclude lower $\sigma$ E+A's at higher
redshift, rather, lower $\sigma$ E+A's simply fall below our strict
selection criteria.  Excellent examples of this point are E+A's
1358--343, 1358--481, and 2053--2081 (Table~\ref{ea_sample}).  Although
fainter than our magnitude limit, all three have high $S/N$ spectra and
{\it measured} velocity dispersions $<110$\kms.

\subsection{Significance of the E+A Phase}\label{importance}
  
\begin{figure}
\plotone{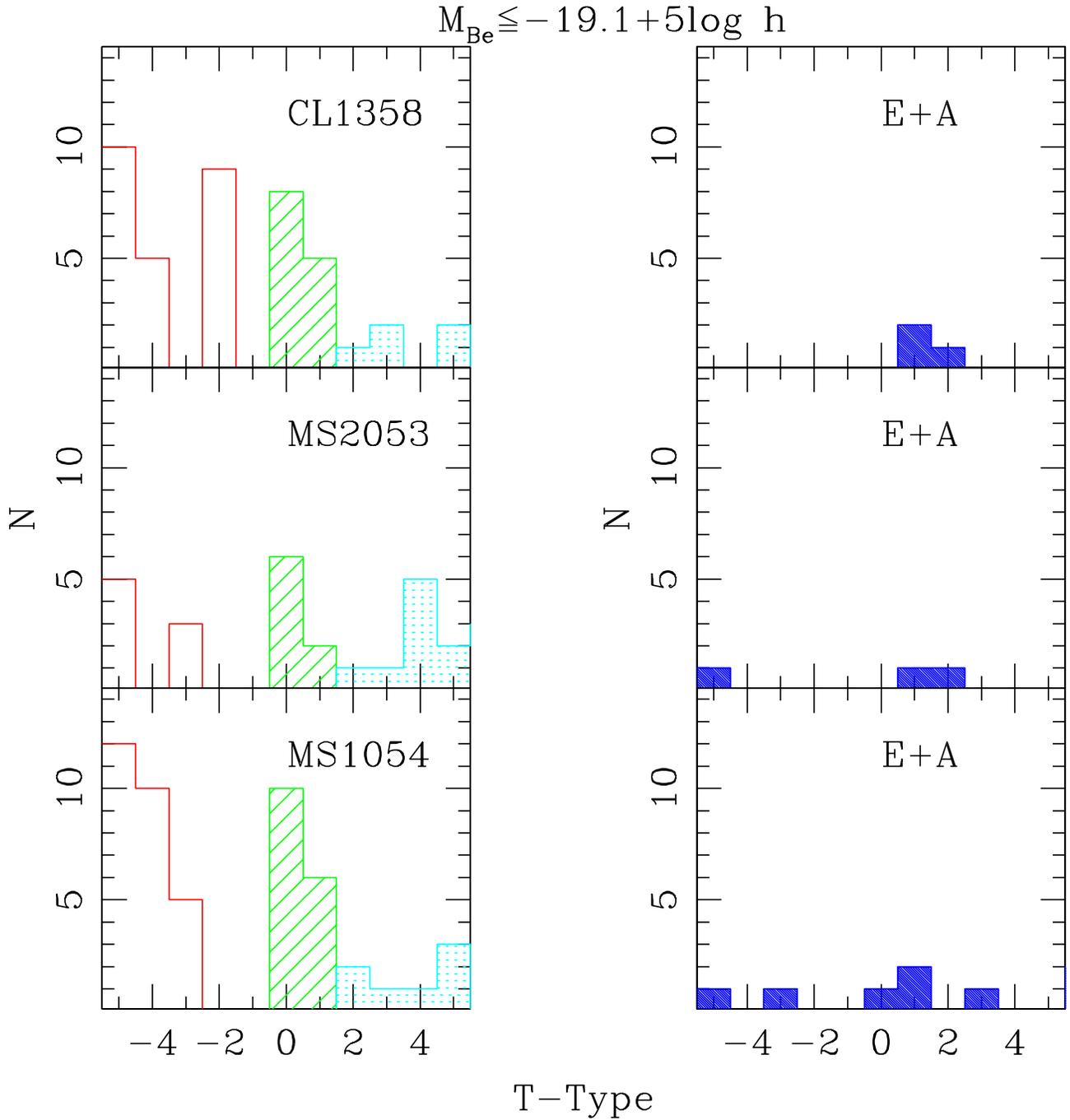}
\caption{Histograms of Hubble type for cluster members above our
magnitude limit.  In the left panels, open regions correspond to E-S0's
($-5\leq T\leq-1$), hatched regions to S0/a-Sa's ($0\leq T\leq1$), and
dotted regions to spirals ($T\geq2$).  The respective E+A
distributions of the three clusters are shown in the right panels; two
of the E+A's in MS1054 are considered mergers (not shown).  The
E+A galaxies span the range of morphological type but tend to be
systems with disks.
\label{morph_hist}}
\end{figure}

Results from our study show that the E+A phase is not limited to a
small fraction of predominately late-type members, but that many
massive early-types also undergo this phase.  To determine if a
significant fraction of cluster early-types have had an E+A phase, we
consider MS1054 ($z=0.83$, $t_{age}\sim 6.5$ Gyr, $H_0=70$ \kms
Mpc$^{-1}$).  In this cluster, $\sim8$\% of the E-S0 members are also
E+A's (see Fig.~\ref{morph_hist}).  To estimate a lower limit on the
number of cluster E-S0's that have undergone an E+A phase by
$z\sim0.8$, we assume the E+A phase can occur at $z\lesssim2.5$
($t_{age}\sim 2.6$ Gyr) and is visible for $\sim1$ Gyr.  In this case,
more than 30\% of the E-S0's in MS1054 have had an E+A phase, with the
restriction that we know of no E+A's with internal velocity dispersions
$>250$\kms.

However, the true number of early-types that have had an E+A phase can
be easily 100\% if we also consider 1) an increasing E+A fraction with
redshift; 2) an increasing characteristic E+A mass with redshift; and
3) the conversion of spirals into early-types.  For example, if the
fraction of early-type E+A's increases to 15\% at $z>0.8$, the total
number of cluster E-S0's that have undergone an E+A phase in MS1054
increases to $\sim70$\%.  As for the observed restriction that only
E+A's with $\sigma<250$\kms~exist at $z\lesssim0.8$, an increasing
characteristic mass may mean there is no upper mass limit on the E+A
phenomenon at higher redshifts; even brightest cluster galaxies (BCG)
may have had an E+A phase.  In addition, morphological transformation
is likely to play a prominent role in increasing the early-type E+A
fraction.  From comparing CL1358 to MS1054 (Fig.~\ref{nsigma_hist}), we
estimate $\sim45$\% of massive ($\sigma>200$\kms) E-S0's are
transformed from spirals; even if only half of these undergo an E+A
phase, the early-type E+A fraction increases by $\sim20$\%.  Any
combination of these factors will only increase the number of E-S0's
that have undergone an E+A phase.

Our estimates on the significance of the E+A phase depend heavily on
how long the E+A phase is visible and whether E+A's evolve into
early-types.  However, even our conservative estimate of $\sim30$\%
establishes the importance of the E+A phase for a non-neglible number
of cluster E-S0's.  If the true fraction is 100\%, as we demonstrate it
can be easily, the E+A phase would be play a critical role in the
transformation of spirals into the early-types that dominate the
cluster population.

To confirm the importance of the E+A phase in the evolution of cluster
galaxies, additional observations in this redshift regime and higher
are needed.  We suspect that more massive cluster members underwent an
E+A phase at $z>0.8$, but whether this also includes the BCG or if
there is an upper mass limit to the E+A phase is unknown.  For a
cluster at $z\sim1.3$, \citet{vandokkum:03} find that the two brightest
members show [OII]$\lambda3727$ emission and enhanced H$\delta$
absorption; the fourth brightest cluster member is an E+A.  Measuring
OII and Balmer indices of BCG's at $0.8<z<1.3$ can establish if BCG's
also undergo an E+A phase, and whether this transition occurs at
$z\sim1$.  Another interesting possibility is to measure indices of
$z>1.5$ radio galaxies as they can be the predecessors of lower
redshift BCG's \citep[e.g.][]{venemans:02}.  With observations at
higher redshift, we can determine if the E+A phase is related to the
bulk of star formation, or if it is simply due to ``frosting'' by a
small fraction of younger stars \citep{trager:00}.

\section{Conclusions}

We estimate the E+A fraction in intermediate redshift clusters and
examine the physical properties of this population using HST/WFPC2
mosaics and extensive ground-based spectroscopy.  Our results are based
on spectral types, galaxy colors, magnitudes, Hubble types, and
quantitative structural parameters for $\sim500$ confirmed members in
three clusters ($z=0.33,~0.58,~\&~0.83$).  We also include measured
internal velocity dispersions for a subset of 120 members, and estimate
dispersions for the rest of the cluster galaxies using the Fundamental
Plane.

We find E+A's make up a non-negligible component of the cluster
population ($\sim7-13$\%) at intermediate redshifts.  They tend to be
bluer than the passive members, and we estimate the E+A phase can
brighten a galaxy by as much as $\Delta M_{Be}\sim1.25$ mag.  Although
most of them are disk-dominated systems, E+A's span the range in
morphological type to include even E-S0 members.  They can be more
luminous than $L^{\ast}$ and can have internal velocity dispersions in
excess of 200\kms.  The majority of these E+A's are not associated with
mergers.

The diverse nature of E+A members indicates they are drawn from a
heterogenous parent population.  The characteristics of their
descendants can be equally varied; even massive early-type members may
have had an E+A phase in their past.  Our study indicates the high
dispersion ($\sigma>200$\kms) E+A's at $z=0.83$ are the logical
progenitors of massive early-types at lower redshift.  

The cluster E+A's are distributed such that the most luminous, high
dispersion ones are in our most distant cluster while only $L\lesssim
L^{\ast}$, low dispersion E+A's exist in our lowest redshift cluster
and Coma \citep{caldwell:96,caldwell:99}.  The trend of increasing
luminosity and internal velocity dispersion with redshift provides
compelling evidence for mass evolution in the cluster E+A population.
This galaxy ``down-sizing'' is similar to the observed decrease in
luminosity of rapidly star-forming field galaxies since $z\sim1$
\citep{cowie:96}.  Comparison of the E+A luminosity, velocity
dispersion, and mass ($\propto r_e\sigma^2$) distributions to the rest
of the cluster members indicate this evolution in the E+A mass
distribution is real.  However, we cannot completely rule out that the
luminous, massive E+A's found at higher redshift are due only to an
increasing E+A fraction.

Using statistical arguments, we find that $\gtrsim30$\% of cluster
E-S0's at $z=0.83$ have had an E+A phase.  We consider this a lower
limit as evolution in the E+A fraction and characteristic mass as well
as the conversion of spirals into early-types can increase the true
fraction to 100\%.  These results show that the E+A phase can play an
important role in the transformation of star-forming galaxies into
early-type members.

Recognizing that our study is based on only three clusters, these
results can only benefit from a similar analysis of other clusters
within this redshift range ($0.3<z<0.8$).  Also needed is a comparative
study of intermediate $z$ field E+A's to determine if field and cluster
E+A's evolve in a similar manner.  Previous studies on the field E+A
fraction at $z>0.2$ disagree \citep[5\% vs.
$\sim0.1$\%;][B99]{hammer:97}, and how the field E+A's evolve in
luminosity and mass has yet to be determined.  Using the considerable
field sample we have acquired in our cluster survey, we will quantify
and characterize the nature of field E+A's in a future paper.

\acknowledgements

We are indebted to M. Balogh for making available the CNOC1 data on
CL1358.  We appreciate the useful comments S. Faber and the referee
provided on this manuscript.  K. Tran thanks the Lorentz Institute for
their generous hospitality during multiple visits.  Support from NASA
HST grants GO-07372-96A, GO-06745-95A, GO-05991-94A, and GO-05989-94A,
and NASA grant NAG5-7697 is gratefully acknowledged.  The authors
thank the entire staff of the W. M. Keck Observatory for their
support, and extend special thanks to those of Hawaiian ancestry on
whose sacred mountain we are privileged to be guests.

\bibliographystyle{/home/vy/aastex/apj}
\bibliography{/home/vy/aastex/tran.bib}


\clearpage

\begin{figure}
\epsscale{1}
\plotone{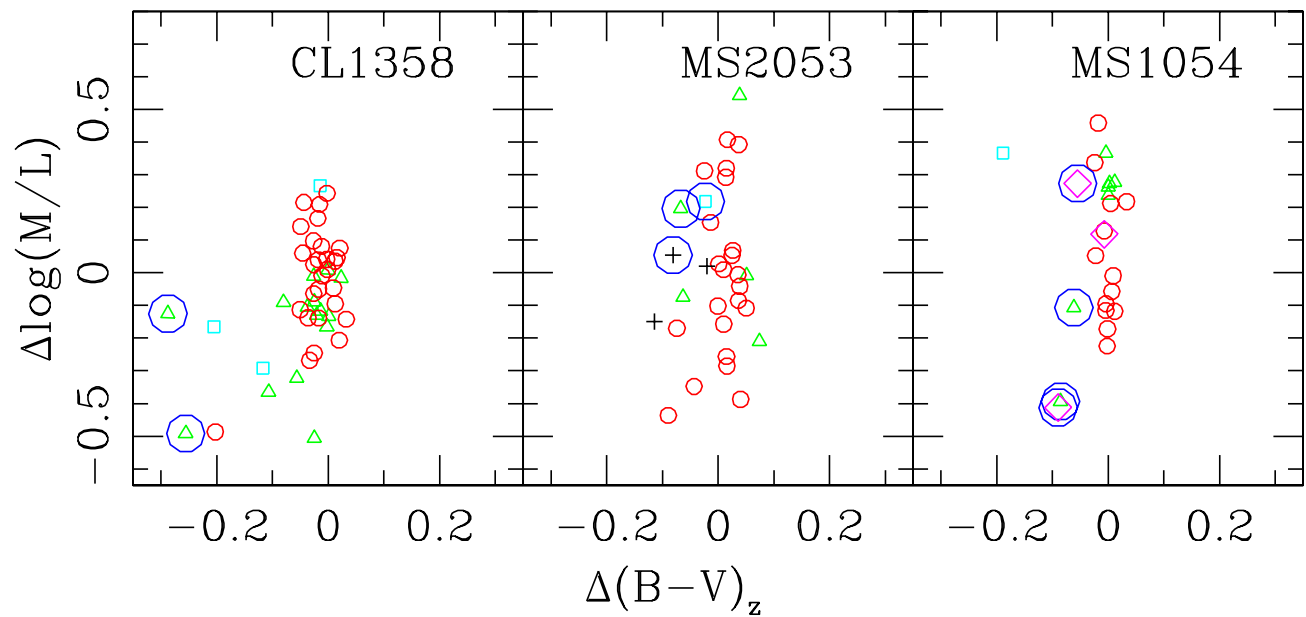}
\caption{The $\log(M/L)$ residuals from the FP plotted against
  residuals from the $(B-V)_z$ color-magnitude relation for members
  with measured internal velocity dispersions; the symbols are as in
  Fig.~\ref{BVz_balmer}.  We estimate cluster galaxies are brightened
  by as much as $\sim1.25$ mag in $M_{Be}$, with a median of $0.25$
  mag.
\label{dlgML_dBV}}
\end{figure}

\clearpage

\begin{deluxetable}{lrrrrrl}
\tablecolumns{5}
\tablewidth{0pc}
\tablecaption{Cluster Sample\label{clusters}}
\tablehead{
\colhead{Cluster} & \colhead{$\bar{z}$} & 
\colhead{$<\sigma>_{CL}$(km s$^{-1}$)} & 
\colhead{$(m-M)$\tablenotemark{a}} & \colhead{Reference}}
\startdata
CL~1358+62& $0.3283\pm0.0003$ & $1027\pm50$ & 40.08 & \citet{fisher:98}\\
MS~2053--04& $0.5840\pm0.0005$ & $870\pm60$ & 41.31 & \citet{tran:02}\\
MS~1054--03& $0.8309\pm0.0006$ & $1130\pm80$ & 42.00 & \citet{tran:99}\\
& & & & \citet{vandokkum:00}\\
& & & & \citet{tran:02}
\enddata
\tablenotetext{a}{Distance modulus determined using $\Omega_M=0.3$,
$\Omega_{\Lambda}=0.7$ cosmology with $H_0=100$\kms~Mpc$^{-1}$, and
corrected for simple fading as determined from the Fundamental Plane
\citep[$\Delta\log(M/L) \propto -0.40z$;][]{vandokkum:98b}.}
\end{deluxetable}

\begin{deluxetable}{lrrr}
\tablecolumns{4}
\tablewidth{0pc}
\tablecaption{Line Strength Index Definitions~\tablenotemark{a} 
\label{indices}}
\tablehead{
\colhead{Index}   & \colhead{Bandpass}      & 
\colhead{Blue Sideband} & \colhead{Red Sideband}}
\startdata
[OII]           & 3716.3--3738.3 & 3696.3--3716.3 & 3738.3--3758.3\\
H$\delta$       & 4083.5--4122.3 & 4017.0--4057.0 & 4153.0--4193.0\\
H$\gamma$       & 4319.8--4363.5 & 4242.0--4282.0 & 4404.0--4444.0\\
H$\beta$        & 4847.9--4876.6 & 4799.0--4839.0 & 3738.3--3758.3\\
\enddata
\tablenotetext{a}{From \citet{fisher:98}.}
\end{deluxetable}

\begin{deluxetable}{lrrrr}
\tablecolumns{5}
\tablewidth{0pc}
\tablecaption{E+A Selection Criteria:  CL1358\tablenotemark{a}
\label{ea_sel}}
\tablehead{
\colhead{Reference} & \colhead{Balmer (\AA)\tablenotemark{b}} & 
\colhead{$\lambda3727$ (\AA)\tablenotemark{b}} & 
\colhead{$N$\tablenotemark{c}} & 
\colhead{Magnitude\tablenotemark{d}}}
\startdata
\citealt{balogh:99} & H$\delta~\geq5$& [OII]$>-5$& 2 & $20.0<m_R<20.9$\\
\citealt{fisher:98} & (H$\delta+$H$\gamma+$H$\beta)/3~\geq4$&
[OII]$>-5$ & 4 & $20.0<m_R<21.4$ \\
\citealt{tran:02} and this paper & (H$\delta+$H$\gamma)/2~\geq4$ & 
[OII]$>-5$ & 4 & $20.0<m_R<21.4$
\enddata
\tablenotetext{a}{Considering only the 98 members in both F98, B99,
and this work.}
\tablenotetext{b}{Equivalent widths determining using the bandpasses
  from Table 2.  Absorption is positive, emission is negative.}
\tablenotetext{c}{Number of E+A galaxies using these selection
  criteria.}
\tablenotetext{d}{Gunn $r$ magnitudes from the CNOC1 survey
  \citep{yee:96}.} 
\end{deluxetable}

\begin{deluxetable}{lrrrrrrrrr}
\tablecolumns{10}
\tablewidth{0pc}
\tablecaption{E+A Sample\label{ea_sample}}
\tablehead{
\colhead{Cluster} & \colhead{ID}    & \colhead{$M_{Be}$\tablenotemark{a}} &
\colhead{$(B-V)_z$\tablenotemark{a}}     & \colhead{T-type\tablenotemark{b}} &
\colhead{$(B/T)$\tablenotemark{c}} & 
\colhead{$r_{1/2}$ (kpc)\tablenotemark{a,c}} &
\colhead{$(R_A)$\tablenotemark{c}} & 
\colhead{$(R_T)$\tablenotemark{c}} &
\colhead{$\sigma$ (\kms)\tablenotemark{d}}}
\startdata
CL 1358+62&  209$^{\ast,}$\tablenotemark{e} & --19.61 &   0.60 &     1 &   0.42 &    3.3 &   0.03 &   0.08 &    107\\
&  328$^{\ast}$ & --19.46 &   0.63 &     1 &   0.45 &    2.3 &   0.05 &   0.09 &     81\\
&  507$^{\ast}$ & --19.17 &   0.68 &     2 &   0.06 &    3.9 &   0.07 &   0.11 &     (52)\\
&  343 & --18.73 &   0.67 &    --2 &   0.91 &    1.7 &   0.01 &   0.06 &     65\\
&   92 & --18.59 &   0.85 &    --2 &   0.75 &    1.6 &   0.02 &   0.09 &    (133)\\
&  226 & --18.29 &   0.68 &     1 &   0.27 &    2.9 &   0.02 &   0.08 &     (37)\\
&  109 & --18.13 &   0.67 &    --2 &   0.41 &    1.5 &   0.01 &   0.07 &     (47)\\
&  243 & --18.09 &   0.41 &     1 &   0.68 &    1.6 &   0.05 &  --0.02 &     (23)\\
&  481 & --18.04 &   0.81 &    --2 &   0.40 &    1.8 &   0.02 &   0.06 &    107\\
& 167 & --18.01 &   0.60 &    --2 &   0.72 &    1.8 &   0.02 &   0.08 &     (33)\\
&  I553 & --17.81 &   0.73 &     5 &   0.00 &    1.7 &   0.03 &   0.24 &     (41)\\
& 1775 & --17.70 &   0.80 &    --4 &   0.36 &    2.7 &   0.04 &   0.12 &     (48)\\
&  190 & --17.66 &   0.82 &     4 &   0.09 &    3.1 &   0.03 &   0.10 &     (51)\\
&  565 & --17.58 &   0.68 &    --2 &   0.74 &    1.5 &   0.00 &   0.11 &     (35)\\
&  420 & --17.37 &   0.62 &     ? &   0.45 &    1.7 &  --0.01 &   0.04 &     (24)\\
& 1414 & --17.29 &   0.78 &     3 &   0.65 &    1.5 &   0.01 &   0.11 &     (49)\\
\\
MS 2053--03& 3549$^{\ast}$ & --19.26 &   0.84 &  --897 \tablenotemark{f} &   0.65 &    2.3 &   0.05 &   0.13 &    154\\
&  416$^{\ast}$ & --19.24 &   0.85 &     1 &   0.34 &    3.5 &   0.04 &   0.10 &    143\\
& 2345$^{\ast}$ & --19.10 &   0.89 &     2 &   0.33 &    2.7 &   0.07 &   0.09 &    151\\
& 1746 & --18.87 &   0.58 &     2 &   0.46 &    1.6 &   0.01 &   0.10 &     (38)\\
& 2081 & --18.67 &   0.79 &     ? &   0.34 &    1.8 &   0.03 &   0.06 &     86\\
& 2265 & --18.36 &   0.77 &     4 &   0.27 &    2.5 &   0.02 &   0.13 &     (39)\\
& 1303 & --18.22 &   0.89 &     3 &   0.77 &    3.9 &  -0.02 &   0.06 &     (61)\\
& 1269 & --18.00 &   0.87 &    --5 &   0.88 &    1.0 &   0.01 &   0.07 &     (90)\\
&  408 & --17.16 &   0.56 &     ? &   0.03 &    1.5 &   0.01 &  0.00 &     (13)\\
\\
MS 1054--03& 5359$^{\ast}$ & --20.56 &   0.86 &     1 &   0.07 &    4.3 &   0.03 &   0.06 &    138\\
& 5840$^{\ast}$ & --20.23 &   0.85 &    99 &   1.00 &    1.2 &   0.02 &   0.06 &    211\\
& 5923$^{\ast}$ & --19.95 &   0.84 &     1 &   0.41 &    3.5 &   0.00 &   0.05 &    (165)\\
& 5534$^{\ast}$ & --19.89 &   0.81 &    --5 &   0.68 &    1.8 &   0.01 &   0.07 &    (152)\\
& 6567$^{\ast}$ & --19.78 &   0.87 &    99 &   0.18 &    3.8 &   0.04 &   0.10 &    201\\
& 2710$^{\ast}$ & --19.77 &   0.88 &     3 &   0.66 &    2.9 &   0.10 &   0.12 &    (140)\\
& 2872$^{\ast}$ & --19.63 &   0.86 &     0 &   0.22 &    1.9 &   0.09 &   0.11 &    179\\
&  987$^{\ast}$ & --19.49 &   0.91 &    --3 &   0.39 &    1.6 &   0.03 &   0.13 &    (221)\\
& 6164 & --19.48 &   0.93 &    --5 &   0.69 &    2.8 &   0.03 &   0.04 &    (174)\\
& 5833 & --18.96 &   0.81 &     0 &   0.25 &    1.8 &   0.02 &   0.09 &     (86)\\
& 5926 & --18.72 &   0.76 &     0 &   0.96 &    1.7 &   0.01 &   0.08 &     (68)\\
& 8001 & --18.70 &   0.85 &    --5 &   0.67 &    0.4 &   0.01 &   0.07 &    (165)\\
& 4390 & --18.66 &   0.93 &    --4 & \nodata &  \nodata & \nodata & \nodata &    \nodata\\
& 5691 & --18.61 &   0.94 &     0 &   0.56 &    1.7 &   0.03 &   0.04 &    (152)\\
& 3356 & --18.46 &   0.79 & ? &   0.92 &    0.9 &   0.04 &   0.10 &     (79)\\
& 6309 & --18.42 &   0.75 &    --5 &   0.45 &    1.7 &  -0.01 &   0.11 &     (56)\\
&  426 & --18.38 &   0.76 & ? &   0.82 &    1.2 &   0.03 &   0.11 &     (65)\\
& 2746 & --18.09 &   0.82 & ? &   0.81 &    1.8 &   0.05 &   0.09 &     (60)\\
& 2467 & --17.86 &   0.81 & ? &   0.96 &    2.7 &   0.06 &   0.15 &     (60)\\
& 4165 & --17.41 &   0.71 & ? & \nodata &  \nodata  & \nodata & \nodata
&    \nodata\\
\\\hline
\\\\\\\\\\
\enddata
\tablenotetext{*}{These are E+A galaxies brighter than our magnitude
limit of $M_{Be}=-19.1$\logh~and with Balmer signal to noise flux ratio
$\geq20$.  }
\tablenotetext{a}{The galaxy identification numbers used here are from
  the SExtractor catalogs of the HST/WFPC2 mosaics.  We assume an
  $H_0=100$\hi\kms Mpc$^{-1}$, $\Omega_M=0.3$, $\Omega_{\Lambda}=0.7$
  cosmology.  Redshifted Johnson magnitudes and colors are converted 
  from F606 and F814 fluxes.  }
\tablenotetext{b}{Hubble types are from \citet{fabricant:00,fabricant:03}:
E-S0 ($-5\leq T \leq-1$), S0/a-Sa ($0\leq T \leq 1$), spirals ($2\leq T
\leq 10$), mergers ($T=99$), and no type ($T=?$).  }
\tablenotetext{c}{Structural parameters are determined from 2D de
  Vaucouleurs+exponential disk fits in the F814W filter
  \citep{tran:02}.  Typical systematic errors for $B/T$, $R_A$, and
$R_T$ are $\sim0.10, 0.02,~\&~0.02$ respectively
\citep{im:01,tran:02,tran:03a}.} 
\tablenotetext{d}{Estimated internal velocity dispersions are noted by ().}
\tablenotetext{e}{This system is not a merger as the two projected
  cluster members have significantly different redshifts ($\Delta
  z>600$ \kms) from the E+A candidate.}
\tablenotetext{f}{Galaxy 2053--3549
has an early-type morphology but there was disagreement between the
three classifiers as to its exact Hubble type \citep{fabricant:03}.}
\end{deluxetable}

\begin{deluxetable}{lrrrrrr}
\tablecolumns{7}
\tablewidth{0pc}
\tablecaption{E+A Fractions\label{eafractions}}
\tablehead{
\colhead{Cluster} & \colhead{$N$\tablenotemark{a}} & 
\colhead{$N_{E+A}$~\tablenotemark{a}} & 
\colhead{$F_{E+A}$~\tablenotemark{a,c}} &
\colhead{$n$\tablenotemark{b}} &
\colhead{$n_{E+A}$~\tablenotemark{b}} &
\colhead{$f_{E+A}$~\tablenotemark{b,c}}    } 
\startdata
CL~1358+62 & 173 & 16 & 9$\pm2$\% & 42 & 3 & 7$\pm4$\% \\
MS~2053--04& 132 & 9  & 7$\pm2$\%  & 30 & 3 & 10$\pm6$\% \\
MS~1054--03& 128 & 20 & 16$\pm3$\% & 61 & 8 & 13$\pm5$\% 
\enddata
\tablenotetext{a}{Using all spectroscopically confirmed members that fall 
  on the HST/WFPC2 mosaic.  We select E+A's as galaxies having 
  [OII]$\lambda3727>-5$\AA~and (H$\delta+$H$\gamma)/2\geq4$\AA.}
\tablenotetext{b}{Considering only members above our magnitude cut of
  $M_{Be}\leq-19.1$\logh.  Robust E+A's are 
  brighter than our magnitude cut and also have H$\delta$ and H$\gamma$
  fluxes with high signal to noise ratios ($\geq20$). }
\tablenotetext{c}{Errors for the E+A fractions are determined by
  assuming the E+A galaxies follow a Poisson distribution. }
\end{deluxetable}

\clearpage
\appendix

\section{Deriving Galaxy Velocity Dispersions with the Fundamental Plane} 

The Fundamental Plane \citep[FP,][]{faber:87,djorgovski:87} is an
empirical relation between galaxy size, surface brightness, and
central velocity dispersion.  It has the form

\begin{equation}\label{fp}
\log r_e = \alpha \log \sigma + \beta \log I_e +\gamma
\end{equation}

\noindent where $r_e$ is the effective radius, $\sigma$ the internal
velocity dispersion, $I_e$ the average surface brightness within $r_e$,
and $\alpha, \beta,$ and $\gamma$ are measured values.  The constants
$\alpha, \beta$, and $\gamma$ are determined by applying Eq.~\ref{fp}
to early-type galaxies where all three observables ($r_e, \sigma,$ and
$I_e$) are known, $e.g.$ \citet{jorgensen:96}. Assuming homology, the
FP implies that mass to light ratios ($M/L$) are a function of galaxy
mass \citep{faber:87}.

\citet{vandokkum:98b} and K00c demonstrated there exists in the FP a
tight correlation with little scatter for cluster early-type galaxies
up to $z\sim0.8$.  Assuming cluster galaxies evolve passively and do
not experience any major mergers, their sizes and internal velocity
dispersions do not evolve.  We can then exploit this correlation and
estimate velocity dispersions from the galaxy's size ($r_e$) and
surface brightness ($I_e$).

We build on these steps by extending our analysis to include galaxies
not observed in K00b.  K00c model the stellar populations of the
cluster members in CL1358 and assume the residuals of the FP and
color-magnitude relation are due to differences in the mean
luminosity-weighted ages (Fig.~\ref{dlgML_dBV}).  By doing so, K00c
find that changes in $M/L$ ratios are strongly correlated with color
deviations alone.  By correcting $M/L$ ratios of later-type (``bluer'')
galaxies, we essentially evolve them onto the color-magnitude relation
defined by the early-types, $i.e.$ these later-types become fainter and
redder.  We then apply Eq.~\ref{fp}, normalized to the early-type
members, to estimate velocity dispersions for the later-type galaxies
as well.

Using a stellar population model of constant metallicity and variable
age, K00c determine the $M/L$ correction for early-type spirals to be

\begin{equation}
\Delta (M/L_V) = 3.22[\Delta(B-V)]^2 + 3.26\Delta(B-V)
\end{equation}

\noindent such that the corrected surface brightness is

\begin{equation}\label{IEcorrect}
I_e' = I_e \times 10^{3.22[\Delta(B-V)]^2 + 3.26\Delta(B-V)}.
\end{equation}

\noindent We measure $r_e$ and $I_e$ by fitting {\it pure de
  Vaucouleurs} surface brightness profiles to all cluster members; for
this profile, $r_e=r_{1/2}$ \citep{binney:87}.  Although E+A's tend to
be disk-dominated systems that are not perfectly fit with a pure
$r^{1/4}$ profile, we do not expect this to bias our results since
\citet{kelson:00a} showed that the combination of surface brightness
and scale-length used in the FP is robust.

From \citet{jorgensen:96}, the measured values of $\alpha$ and $\beta$
are 1.16 and $-0.76$ respectively.  We determine the normalization
$\gamma$ using the cluster E-S0's for which we have observed values of
$\sigma$, $r_e$, and $I_e$.  Velocity dispersions then are estimated
for the other cluster members by correcting their surface brightnesses
via Eq.~\ref{IEcorrect}, and then using $I_e'$ and $r_e$ in
Eq.~\ref{fp} with $\gamma_{E-S0}$.  For cluster members with measured
and estimated velocity dispersions, we find the scatter to be
$\sim20$\%, $\sim30$\%, and $\sim35$\% for CL1358, MS2053, and MS1054
respectively.

\end{document}